\documentclass[aps,pra,twocolumn,superscriptaddress]{revtex4}
\usepackage{amsmath}
\usepackage{amssymb}

\usepackage{amsmath,amssymb}
\usepackage[usenames]{color}
\usepackage{mathbbol}
\usepackage{amssymb}
\usepackage{grffile}
\usepackage{graphicx}
\usepackage{amsmath, amstext, amssymb, amsfonts, amsxtra}
\usepackage{textcomp}
\usepackage{xspace} 
\usepackage[colorlinks]{hyperref}

\usepackage{subfigure}

\DeclareSymbolFontAlphabet{\amsmathbb}{AMSb}

\newcommand{\cop}{\hat{c}^{\phantom{\dagger}}}
\newcommand{\cdop}{\hat{c}^{\dagger}}

\newcommand{\nop}{\hat{n}}

\newcommand{\tr}{{\rm tr}}

\newcommand{\ek}{\epsilon_{k}}
\newcommand{\ket}[1]{| #1 \rangle}
\newcommand{\bra}[1]{\langle #1 |}
\newcommand{\expval}[1]{\langle #1 \rangle}
\newcommand{\braOpket}[3]{\langle #1 | #2 | #3 \rangle}

\newcommand{\abs}[1]{|#1|}
\newcommand{\Um}[1]{\"#1}

\newcommand{\bonnpi}{Physikalisches Institut, University of Bonn, Nussallee 12, 53115 Bonn, Germany}
\newcommand{\unbc}{Department of Physics, University of Northern British Columbia, Prince George, British Columbia V2N 4Z9, Canada}
\newcommand{\strathclyde}{Department of Physics and SUPA, University of Strathclyde, Glasgow G4 0NG, United Kingdom}

\begin{document}

\title{Radio-frequency driving of an attractive Fermi gas in a one-dimensional optical lattice}

\author{Johannes Kombe}
\affiliation{\strathclyde} 
\author{Michael K\Um{o}hl}
\affiliation{\bonnpi}
\author{Corinna Kollath}
\affiliation{\bonnpi}
\author{Jean-S\'ebastien Bernier}
\affiliation{\unbc}

\begin{abstract}
  We investigate the response to radio-frequency driving of an ultracold gas of attractively interacting fermions in a one-dimensional optical lattice. We study the system dynamics by monitoring the driving-induced population transfer to a third state, and the evolution of the momentum density and pair distributions. Depending on the frequency of the radio-frequency field, two different dynamical regimes emerge when considering the evolution of the third level population. One regime exhibits (off)resonant many-body oscillations reminiscent of Rabi oscillations in a discrete two-level system, while the other displays a strong linear rise. Within this second regime, we connect, via linear response theory, the extracted transfer rate to the system single-particle spectral function, and infer the nature of the excitations from Bethe ansatz calculations. In addition, we show that this radio-frequency technique can be employed to gain insights into this many-body system coupling mechanism away from equilibrium. This is done by monitoring the momentum density redistributions and the evolution of the pair correlations during the drive. Capturing such non-equilibrium physics goes beyond a linear response treatment, and is achieved here by conducting time-dependent matrix product state simulations.
\end{abstract}

\date{\today}

\maketitle

\section{Introduction}
In recent years, significant experimental efforts have been devoted to dynamically generate complex states and study their evolution. Ultrafast optical pulses were used to photo-induce phase transitions in strongly interacting solid state materials~\cite{BasovHaule2011,FaustiCavalleri2011,Orenstein2012,ZhangAveritt2014,GiannettiMihailovic2016,MitranoCavalleri2016} and similar successes were reported for ultracold atoms using time-dependent electromagnetic fields~\cite{BlochZwerger2008,PolkovnikovVengalattore2011,BehrleKoehl2018}. However, uncovering the mechanisms underlying the non-equilibrium dynamics of strongly correlated matter is still a subject of active research.

Radio-frequency (rf) spectroscopy has established itself as a powerful experimental probe to study the equilibrium properties of ultracold atomic gases~\cite{HaussmannZwerger2009,Torma2016}. Based on the idea of coherent transfer between different internal states of the atom (e.g. different hyperfine levels of the electronic ground state manifold), rf spectroscopy has been successfully applied to measure  (unitarity-limited) `clock' shifts around a Feshbach resonance~\cite{RegalJin2003a,GuptaKetterle2003,BaymZwierlein2007}, and study pairing and molecule formation on the BEC side of the Feshbach resonance~\cite{TormaZoller2000,RegalJin2003b,KinnunenTorma2004a,BartensteinJulienne2005}, as well as the excitation spectrum and underlying pairing gap of interacting Fermi gases~\cite{ChinGrimm2004,KinnunenTorma2004b,SchunckKetterle2008}. More recently, a spatially resolved rf technique has been developed~\cite{ShinKetterle2007}, circumventing complications of density inhomogeneities in harmonically trapped gases, while the momentum-resolved rf spectroscopy introduced in~\cite{StewartJin2008} gives direct access to the spectral function.

Commonly, the obtained rf spectra are interpreted within the framework of linear response. There, the rf field is assumed to only weakly perturb the system, implying that the observed response is that of the unperturbed, equilibrium system. In this limit, the transfer rate is related to a response function. In the absence of final state interactions, the expression simplifies and the transferred particle rate can be shown to be directly related to the single-particle spectral function~\cite{HeLevin2005,PunkZwerger2007,BerthodGiamarchi2015,Torma2016}, as observed in~\cite{StewartJin2008}.

While final state interactions can be neglected in certain systems due to a suitable arrangement of the Feshbach resonance of the Zeeman levels~\cite{SchunckKetterle2008}, generally this is not the case. The spectra are changed both quantitatively and qualitatively, which complicates their interpretation significantly~\cite{YuBaym2006,PeraliStrinati2008,BasuMueller2008,PieriStrinati2009,PieriGrimm2011}.

We use a combination of time-dependent matrix product state simulations (t-MPS)~\cite{WhiteFeiguin2004,DaleyVidal2004,Schollwoeck2011} and analytic techniques, to study the rf response of a half-filled attractive Hubbard model. We investigate the system dynamics by monitoring the driving-induced population transfer to a third state, and the evolution of the momentum density and superconducting pair distributions resulting from this perturbation. Considering the evolution of the population in the third level, we observe two distinct dynamical regimes. One is reminiscent of Rabi oscillations in a driven two-level system, while the other one displays a resonant behaviour, indicating the rf coupling to a continuous band of excitations. We interpret some of the features occurring at weak driving by comparing our numerical results to analytical calculations based on linear response theory combined with Bethe ansatz calculations. From this analysis, we find that certain excitations occurring within the spin-charge continuum of the attractive Hubbard model strongly couple to the rf drive. These excitations can be experimentally detected by monitoring the momentum-resolved density of the final state. Moreover, from the evolution of the momentum density distribution for all three states and the superconducting pair correlations, we can gain insights into the coupling mechanisms at work within this attractively interacting many-body system. Monitoring the non-equilibrium behaviour for these quantities is only possible within our numerical simulations as a linear response treatment would not succeed in capturing their full dynamics.

The paper is organized as follows. We begin in section~\ref{sec:methods_and_modelling} by describing the theoretical model and the analytic techniques used to study the rf response of the interacting Fermi gas. In section~\ref{sec:mps} we introduce the numerical method used to simulate the many-body problem. Section~\ref{sec:weak_attraction} introduces the central observables, and presents the results obtained for a weakly interacting initial state, while section~\ref{sec:strong_attraction} contrasts this to the response for a strongly interacting system. Finally, we conclude with a summary in section~\ref{sec:conclusion}.

\section{Radiofrequency driving of attractively interacting fermions}
\label{sec:methods_and_modelling}
We study the dynamic response to a rf field of attractively interacting fermions prepared in two internal levels and confined to a one-dimensional lattice geometry. In the following sections, we describe how we model the Fermi gas and the rf drive, and briefly discuss the limit of vanishing interactions in the initial state.

\subsection{Attractive Hubbard model}
The Fermi gas is initially prepared in two attractively interacting internal states and is confined to a one-dimensional optical lattice. For sufficiently deep lattice potentials, the unperturbed Hamiltonian describing this system can be approximated by the Hubbard Hamiltonian, 

\begin{equation}
  H_{0} = - J \sum_{i = 1}^{L-1} (\cdop_{i,\sigma}\cop_{i+1,\sigma} + \text{h.c.}) + U \sum_{i=1}^{L} \nop_{i,1}\nop_{i,2} ~ ,
  \label{eq:H0} 
\end{equation} 
where $\hat{c}^{(\dagger)}_{i,\sigma}$ are the fermionic annihilation (creation) operators of the internal level $\sigma = \{1,2\}$ on site i, and $\nop_{i,\sigma}$ is the corresponding number operator. $J$ denotes the hopping amplitude of the fermions, $U < 0$ the attractive on-site interaction, and $L$ the number of lattice sites. At half-filling the ground state of the Hubbard model undergoes a quantum phase transition at $U = 0$, where the system is a Mott insulator for all $U > 0$ and metallic for $U \leq 0$~\cite{EsslerKorepin2005} (and references therein). For half-filling and $U<0$, conformal field theory and bosonization predict superconducting (SC) and charge density wave (CDW) correlations both to decay algebraically as $|d|^{-\nu}$ ($\nu = 1$), where $d$ is the distance between the sites on which the correlations are taken. In contrast, spin density wave (SDW) correlations are exponentially suppressed with the distance $d$ (and vice versa for $U>0$). Below half-filling, SC and CDW decay algebraically with distance, but the SC correlations dominate for attractive interactions~\cite{BogolyubovKorepin1990,EsslerKorepin2005}.

\subsection{Identifying the excitations above the ground state}
The one-dimensional Fermi-Hubbard model $H_{0}$, Eq.~\ref{eq:H0}, is integrable and exactly solvable using Bethe ansatz techniques~\cite{LiebWu1968,EsslerKorepin2005}. In this section, we briefly summarize the derivation of the Bethe ansatz for the attractive Hubbard model with an emphasis on its excitation spectrum, as this will be important in the analysis of the weak rf drive.

Within the Bethe ansatz, the obtained eigenfunctions are determined by two sets of quantum numbers $\{q_{j}\}$ and $\{\lambda_{m}\}$, known as charge momenta and spin rapidities, respectively. Using these quantum numbers, the energy $E$ and momenta $P$ of the elementary excitations can be expressed as 

\begin{equation}
  E = -2J \sum_{j=1}^{N_{1}+N_{2}} \cos(q_{j}) ~ , ~ P = \left( \sum_{j=1}^{N_{1}+N_{2}} q_{j} \right) \text{mod} ~ 2\pi ~ ,
  \label{eq:Bethe_EnergyMomentum}
\end{equation}
where $N_{\sigma}$ is the total number of fermions in internal level $\sigma$. In general (particularly in the attractive model), these parameters are complex and satisfy the Lieb-Wu equations, 

\begin{align}
  e^{iq_{j}L} =& \prod_{m=1}^{N_{2}} \frac{\lambda_{m} - \sin(q_{j}) - iU/4}{\lambda_{m} - \sin(q_{j}) + iU/4} \nonumber \\
  \prod_{j=1}^{N_{1}+N_{2}} \frac{\lambda_{m} - \sin(q_{j}) - iU/4}{\lambda_{m} - \sin(q_{j}) + iU/4} =& 
  \prod_{n \neq m}^{N_{2}} \frac{\lambda_{m} - \lambda_{n} - iU/2}{\lambda_{m} - \lambda_{n} + iU/2} ~ ,
  \label{eq:LiebWu}
\end{align}
where $j = \{1, \ldots, N_{1}+N_{2}\}$ in the first and $m = \{1, \ldots, N_{2}\}$ in the second line of Eq.~\ref{eq:LiebWu}. At half-filling (and zero magnetic field) in the thermodynamic limit, one obtains decoupled closed-form equations for the elementary spin- and charge-wave excitations. They are given for the elementary spin excitations by

\begin{eqnarray}
  \epsilon_{sw}(q) &=& \frac{|U|}{2} - 2J\cos(q) + \nonumber \\
                      && 2\int_{0}^{\infty} \frac{d\omega}{\omega} \frac{J_{1}(\omega) \cos(\omega\sin(q))e^{-\omega|U|/4}}{\cosh(\omega U/4)} \nonumber \\
  p_{sw}(q)  &=& q -  \int_{0}^{\infty} \frac{d\omega}{\omega} 
  \frac{J_{0}(\omega) \cos(\omega\sin(q))e^{-\omega|U|/4}}{\cosh(\omega U/4)} \nonumber \\
  \label{eq:s_excitations}
\end{eqnarray}
and for the elementary charge excitations by

\begin{eqnarray}
  \epsilon_{cw}(\lambda) &=& 2\int_{0}^{\infty} \frac{d\omega}{\omega} \frac{J_{1}(\omega) \cos(\omega\lambda)}{\cosh(\omega U/4)} \nonumber \\
  p^{p}_{cw}(\lambda) &=& \pi -  \int_{0}^{\infty} \frac{d\omega}{\omega} \frac{J_{0}(\omega) \sin(\omega\lambda)}{\cosh(\omega U/4)} \nonumber \\
  						&=& \pi - p^{h}_{cw} ~,
  \label{eq:c_excitations}
\end{eqnarray}
where $J_{n}(\omega)$ are Bessel functions and $p^{p/h}_{cw}$ denotes the momentum of a charge-wave excitation of particle (p) or hole (h) character, which are also referred to as antiholon and holon~\cite{EsslerKorepin2005}.

\begin{figure} 
  \includegraphics[width=1.0\columnwidth]{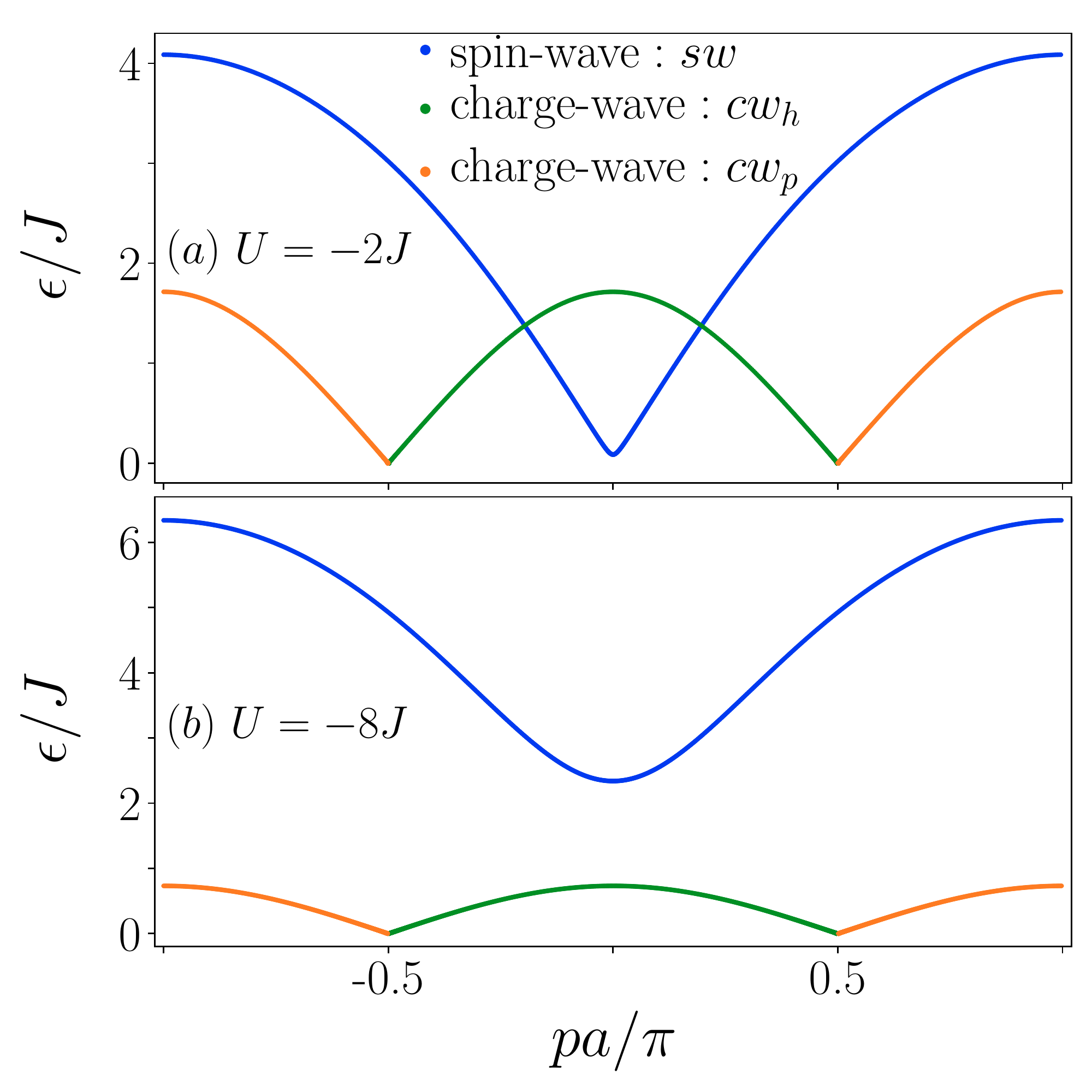}
  \caption{Elementary excitations of the attractive Fermi-Hubbard model from the Bethe ansatz Eq.~\ref{eq:s_excitations} and Eq.~\ref{eq:c_excitations} for $U=-2J$ (a) and
          $U=-8J$ (b).}
  \label{fig:BetheExcitations_SingleParticle}
\end{figure}

\begin{figure} 
  \includegraphics[width=1.0\columnwidth]{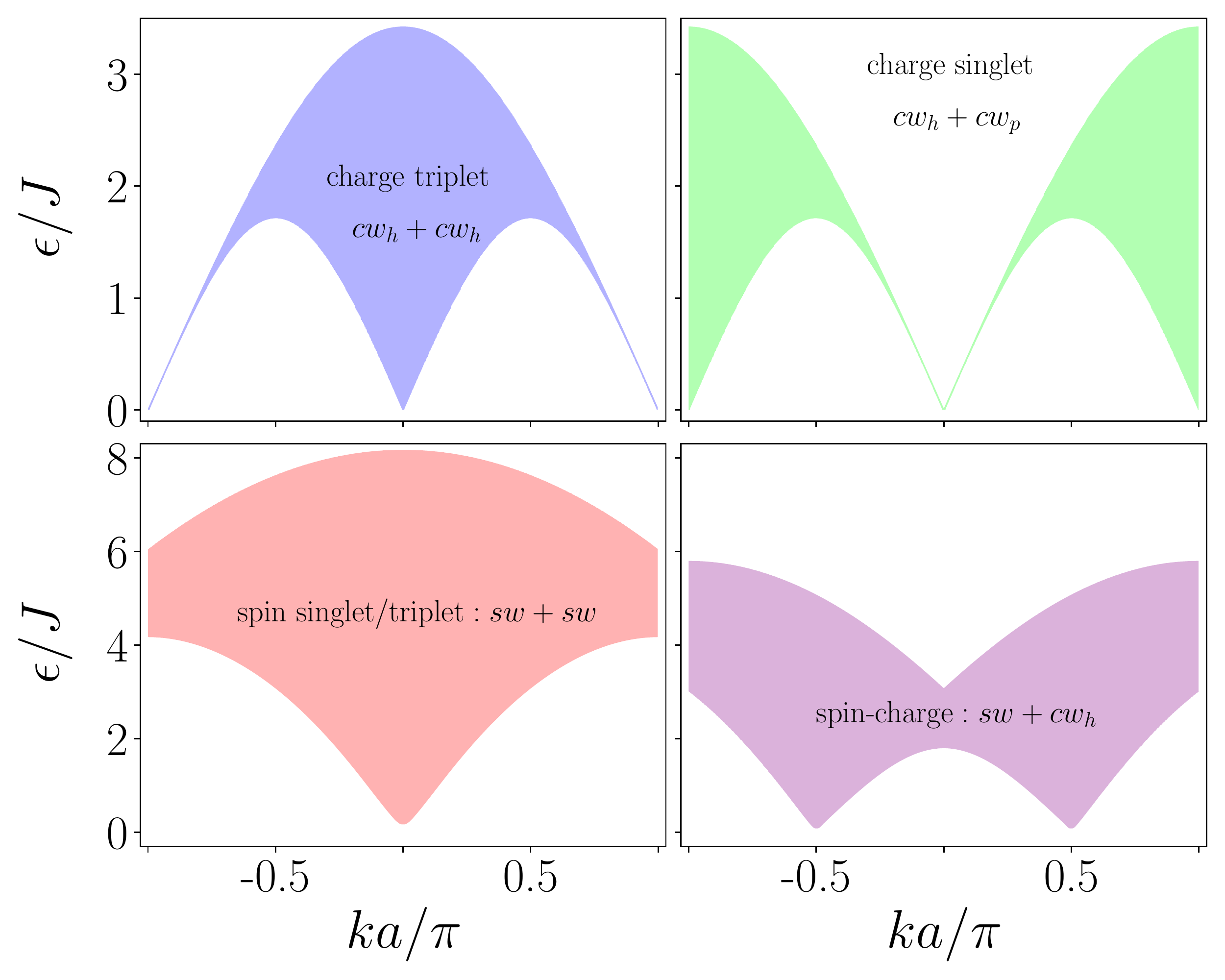}
  \caption{Excitation continua of the attractive Fermi-Hubbard model for $U=-2J$, constructed from Fig.~\ref{fig:BetheExcitations_SingleParticle} where e.g.
          $k = p_{sw} + p^{h}_{cw}$ for the spin-charge continuum  (bottom left).}
  \label{fig:BetheExcitations_TwoParticle_weakU}
\end{figure}

\begin{figure} 
  \includegraphics[width=1.0\columnwidth]{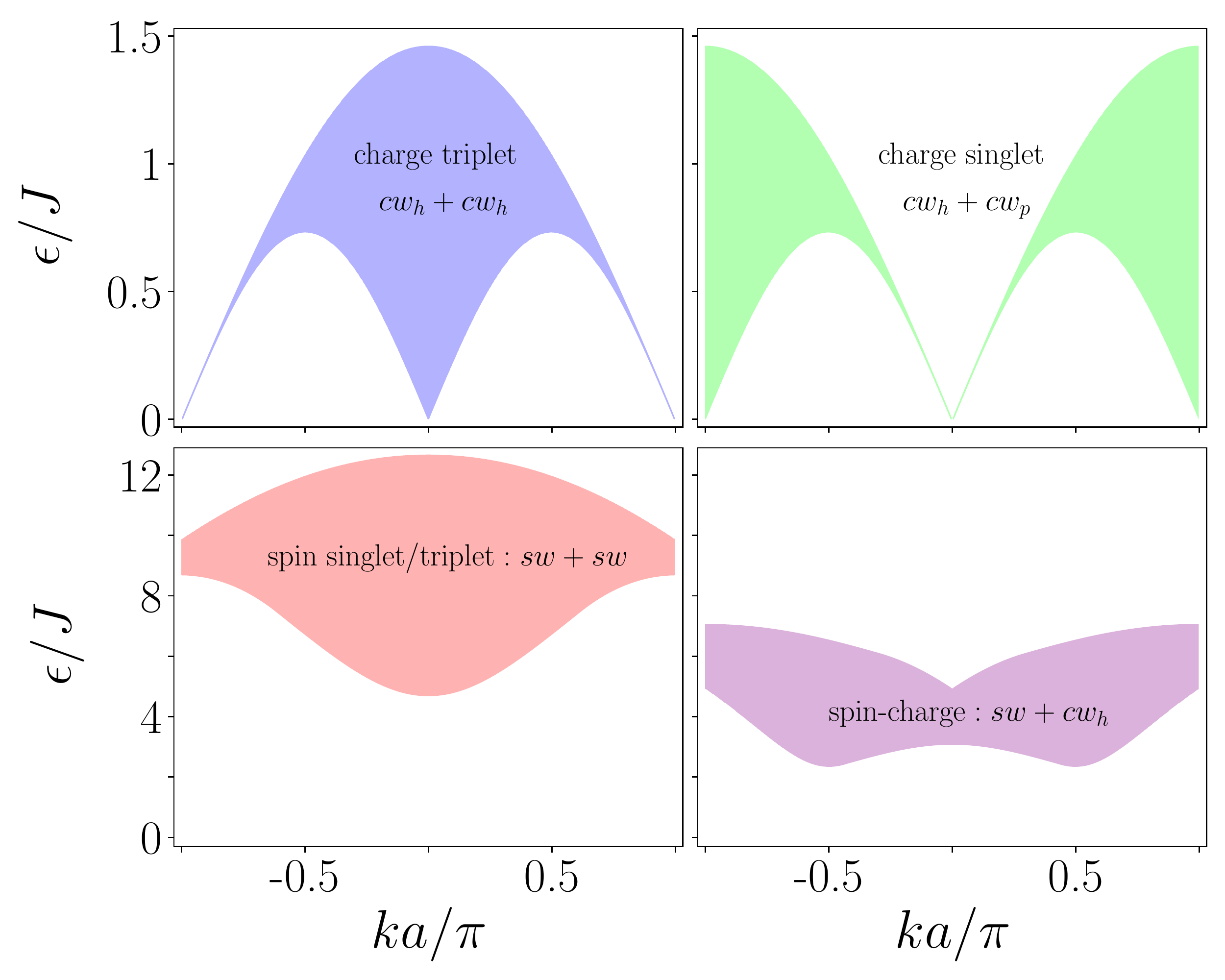}
  \caption{Excitation continua of the attractive Fermi-Hubbard model for $U=-8J$, constructed from Fig.~\ref{fig:BetheExcitations_SingleParticle} where e.g.
          $k = p_{sw} + p^{h}_{cw}$ for the spin-charge continuum  (bottom left).}
  \label{fig:BetheExcitations_TwoParticle_strongU}
\end{figure}

Fig.~\ref{fig:BetheExcitations_SingleParticle} shows the elementary excitations of the Hubbard model at half-filling for weak (a) and strong (b) attraction. We note that the spin-wave is gapped, while the charge-wave remains gapless. However it is important to emphasize that the physical excitations reached by the rf drive need to be constructed from even combinations of elementary excitations. The spin-charge continuum for instance is constructed from two elementary excitations as $\epsilon_{sc} = \epsilon_{sw}(q) + \epsilon_{cw}(\lambda)$, where $k = p_{sw}(q) + p^{h}_{cw}(\lambda)$~\cite{EsslerKorepin1994}. The various excitation continua for weak and strong interaction are shown in Fig.~\ref{fig:BetheExcitations_TwoParticle_weakU} and Fig.~\ref{fig:BetheExcitations_TwoParticle_strongU} respectively. We find the charge singlet and triplet excitations to be gapless (upper panels), while the spin singlet, triplet and the spin-charge continua remain gapped (lower panels). In the following we will focus on the spin-charge continuum as this is relevant for the here investigated rf driving scheme. 

\subsection{Modelling the radiofrequency driving}
\begin{figure} 
  \includegraphics[width=1.0\columnwidth]{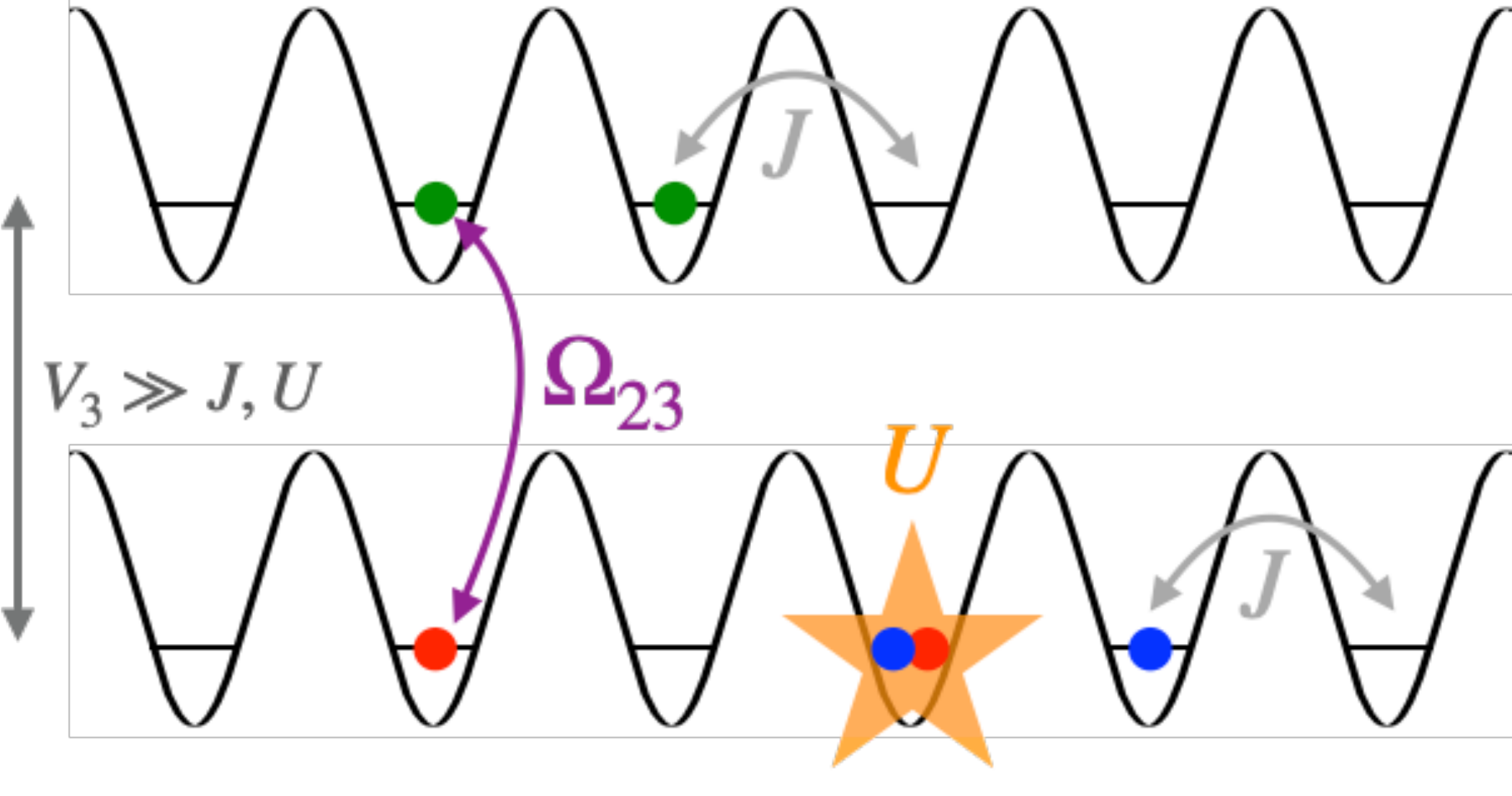}
  \caption{Sketch of the underlying Fermi Hubbard model with the rf coupling of the different internal (hyperfine) states $\sigma = \{1,2,3\}$ (here depicted in blue, red and
          green respectively). The hopping amplitude $J$ is taken to be the same in the lower and upper bands, and the hyperfine splitting to the final state is denoted by
          $V_{3}$.}
  \label{fig:sketch_model}
\end{figure}

The rf field induces transitions between different internal states of the atoms. Here we assume that the rf field induces mainly a transition from the internal state $\sigma=2$ to a third state $\sigma=3$ as sketched in Fig.~\ref{fig:sketch_model}.

In this situation the rf coupling can be modelled by the term 

\begin{eqnarray}
  H'(t) &=& \overbrace{\hbar \Omega_{23} \cos(\omega_{\text{rf}}t)}^{\gamma(t)} \sum_{i=1}^{L} (\cdop_{i,3} \cop_{i,2} + \text{h.c.}) \nonumber \\
        &=& \gamma(t) \sum_{m=1}^{L} (\cdop_{k_{m},3} \cop_{k_{m},2} + \text{h.c.}) ~ ,
  \label{eq:Hprime}
\end{eqnarray}
where $\Omega_{23}$ is the Rabi frequency of the transition (related to the dipole matrix element), $\omega_{\text{rf}}$ the frequency of the rf field, and $k_{m}a = \frac{m\pi}{L+1}$ ($m = 1,\ldots,L$) the momentum of the particle. In the last line of the previous equation, we have used the Fourier transform for open boundary conditions for numerical convenience as explained in section~\ref{sec:mps}. The representation of the coupling in momentum space makes explicit that the rf field drives vertical transition in momentum space which means that no momentum change is transferred by the long wavelength rf pulse.

Additionally, we also need to consider the Hamiltonian of the third level here taken to be free,

\begin{equation}
  H_{3} = - J \sum_{i = 1}^{L-1} (\cdop_{i,3}\cop_{i+1,3} + \text{h.c.}) + V_{3} \sum_{i=1}^{L} \nop_{i,3} ~ .
  \label{eq:H3}
\end{equation}
This neglects the final state interaction which in many experiments can be dominant. However for example in $^{40}\textrm{K}$ final state interactions are small~\cite{StewartJin2008,FeldKoehl2011}, and we are mainly interested in the dynamics induced by the rf driving. The energetic splitting $V_{3}$ between the state $\ket{2}$ and $\ket{3}$ is usually much larger than the kinetic and interaction energy scales, i.e. $V_{3} \gg J, U$. 

Finally, the full model is given by $H(t) = H_{0} + H_{3} + H'(t)$. We note that, since $[N_{1}, H(t)] = 0$, the a priori different hyperfine levels can be brought on top of each other through a unitary transformation. This has been used in order to set the same chemical potential for the level $1$ and $2$, i.e. $V_{1} = V_{2} = 0$.

\subsection{Non-interacting system}
\label{sec:noninteracting_system}
In this subsection, we briefly describe the response of the system in the absence of interaction, i.e. $U=0$. In this case, the Hamiltonian is diagonal in momentum space, the individual momenta $k_{m}$ fully decouple, and one can view the system as a series of three-level quantum systems, subject to a periodic drive,

\begin{equation}
  H(t) = \sum_{k} \Psi_{k}^{\dagger} \begin{pmatrix} \ek & 0 & 0 \\ 0 & \ek & \gamma(t) \\ 0 & \gamma(t) & \ek + V_{3} \end{pmatrix} \Psi_{k} ~,
\end{equation}
where $\ek = -2J\cos(k)$ and $\Psi^{\dagger}_{k} = (\cdop_{k,1}, \cdop_{k,2}, \cdop_{k,3})$. Level $\ket{1}$ is fully decoupled, and the non-trivial dynamics takes place in the two-dimensional \{$\ket{k,2}$, $\ket{k,3}$\} subspace. 

\begin{figure} 
  \includegraphics[width=1.0\columnwidth]{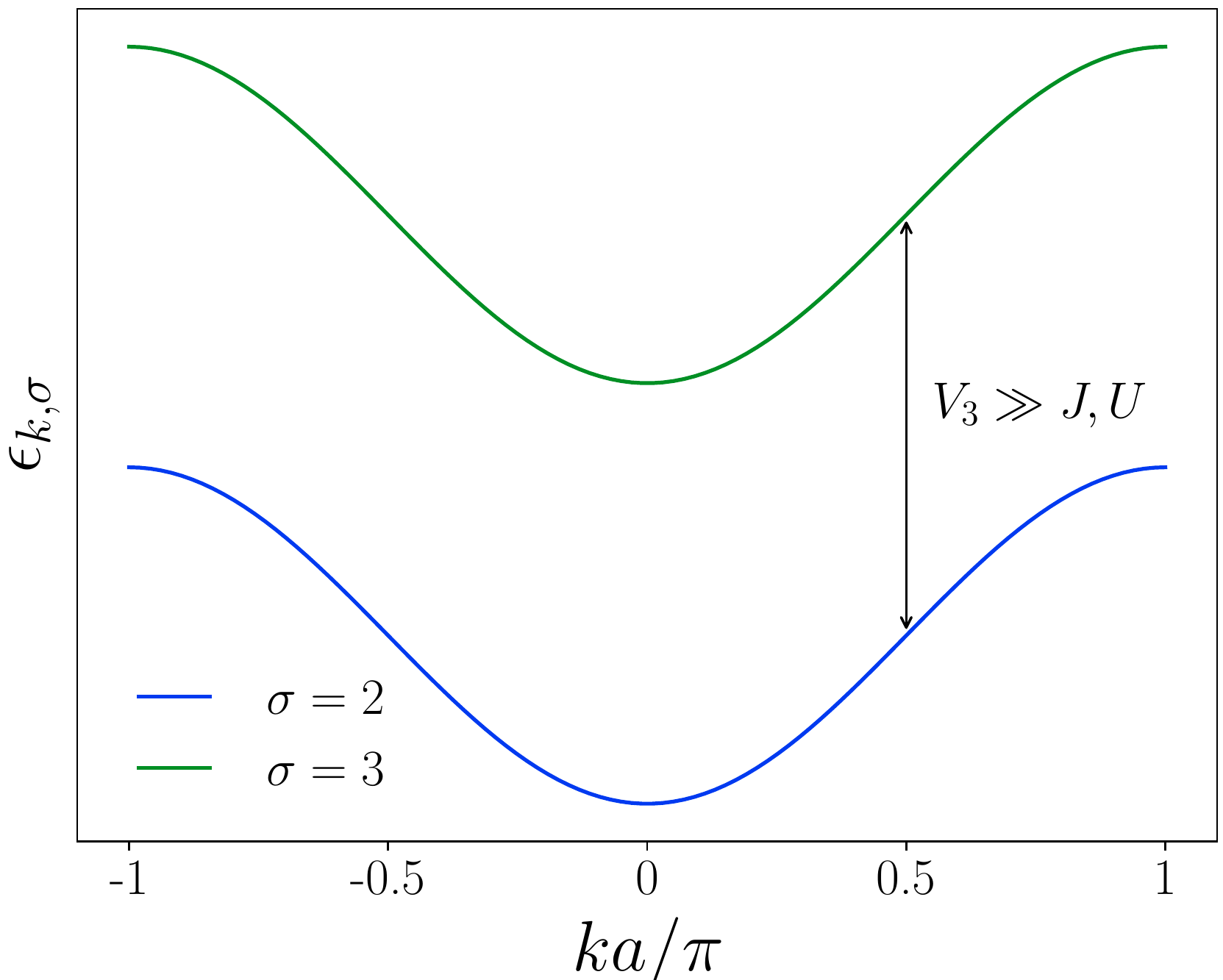}
  \caption{The lower and upper bands of the non-interacting system in the $\ket{23}$ manifold. The two free bands are separated by the hyperfine splitting $V_{3} \gg J,U$.}
  \label{fig:noninteracting_bands}
\end{figure}

Since the wavelengths of rf fields are very long ($\lambda \sim 1m$), there is negligible momentum transfer and the transition is `vertical' in momentum space, as depicted in Fig.~\ref{fig:noninteracting_bands}. 

The effective Hamiltonian of this two-level system (neglecting constant shifts) takes the form $H_{23} = -\frac{V_{3}}{2}\sigma_{z} + \gamma(t) \sigma_{x}$, which we recognize as the Hamiltonian describing a two-level atom driven by a laser field. Within the rotating-wave approximation, the dynamics can be solved analytically. The drive induces (off)resonant Rabi oscillations given by

\begin{equation}
  \expval{\nop_{k,3}(t)} = \frac{\Omega_{23}^{2}}{\Omega_{\text{eff}}^{2}} \sin^{2}\left(\frac{1}{2} \Omega_{\text{eff}}t\right) ~ ,
  \label{eq:Rabioscillation}
\end{equation}
where $\Omega_{\text{eff}} = \sqrt{\Omega_{23}^{2} + \delta^{2}}$ is the generalized, effective Rabi frequency and $\hbar\delta = \hbar\omega_{\text{rf}} - V_{3}$ the detuning of the rf field from the bare 2-3 transition. For finite detunings the Rabi oscillations become faster, albeit with a reduced amplitude. The overall amplitude of the oscillations has a Lorentzian dependence on the detuning, where its width is given by the bare Rabi frequency $\Omega_{23}$ of the problem. Note, that here due to the assumption that the dispersion in level $2$ and $3$ are the same, the oscillation frequency does not depend on the momentum. This will change if the atoms in level 3 feel a different optical lattice potential resulting in a dephasing of the oscillations. 

If the interaction is non-zero, this will introduce a coupling between different momentum sectors and the dynamics will become much more complex. In the following, we will analyze the ensuing dynamics, both for weak ($U=-2J$) and strong ($U=-8J$) interactions, with a combination of analytical response calculations (section~\ref{sec:linear_response}) and numerical simulations (section~\ref{sec:mps}).

\subsection{Linear response theory}
\label{sec:linear_response}
In this section we discuss the linear response theory often employed in order to analyze the response of a system to an rf drive. The validity of this approach depends on two basic requirements. Firstly, the external drive must be weak, so that the system is only weakly perturbed. Secondly, the perturbation must couple the initial state to a continuous band of final states. The drive to the individual levels of the final band quickly dephase, thus one sums over transition probabilities instead of quantum amplitudes, and Rabi oscillations give way to a linear increase in the upper level's population.

For a weak probe, $H'(t)$, the response of an observable $\mathcal{O}$ can often be related to equilibrium expectation values in the unperturbed model. We concentrate in this section on the momentum occupation $\mathcal{O}=\nop_{k,3}$. The expectation value can be computed as a perturbation series in the driving amplitude, which to first-order reads,

\begin{equation}
  \expval{\nop_{k,3}(t)}^{(1)} = - \frac{i}{\hbar} \int_{-\infty}^{t} dt' \expval{[\nop_{k,3}(t), H'(t')]} = 0 ~,
\end{equation}
since level $\ket{3}$ is initially empty. We obtain the first non-zero contribution at second-order,

\begin{align}
  \expval{\nop_{k,3}(t)}^{(2)} = & \left(- \frac{i}{\hbar}\right)^{2} \int_{-\infty}^{t} dt_{1} 
                                    \int_{-\infty}^{t_{1}} dt_{2} \nonumber \\ 
                            & \times \expval{\Big[ [\nop_{k,3}(t), H'(t_{1})], H'(t_{2}) \Big]} ~.
\end{align}
Equivalently, in the following we will look at the first-order response of the transfer rate. After an initial, transient period, the population of the upper level enters a linear regime for a sufficiently weak Rabi frequencies $\Omega_{23}$, and we extract this slope, both for $N_{3}(t)$ and $\nop_{k,3}(t)$. Within linear response theory, the rate of particles transferred from $\ket{2}$ to $\ket{3}$ can be related to the single-particle spectral function $A(k,\omega)$, as 

\begin{eqnarray}
  \expval{\dot{\nop}_{k,3}}^{(1)} &=& - \frac{i}{\hbar} \int_{-\infty}^{t} dt' \expval{[\dot{\nop}_{k,3}(t), H'(t')]} \nonumber \\
  &\sim&  \frac{\pi\Omega_{23}^{2}}{2} \Big[ A(k,\omega_{\text{rf}}) + A(k,-\omega_{\text{rf}}) \Big] ~ ,
  \label{eq:linear_response}
\end{eqnarray}
where in the last line we have neglected fast oscillating terms and only retained the constant background contribution to the slope. Here $A(k,\omega_{\text{rf}}) = \sum_{n} \abs{\braOpket{n}{\cop_{k,2}}{GS}}^{2} \delta(\omega_{n} + \omega_{k,3} - \omega_{0} - \omega_{\text{rf}})$, $\ket{GS} = \ket{\psi(0)}$ is the initial ground state of the system, $\hbar\omega_0$ its energy, and $\ket{n}, \hbar\omega_{n}$ the eigenstates and eigenenergies of $H_{0}$ respectively~\cite{PunkZwerger2007}. The upper level $\ket{3}$ is modelled as a free band, $\hbar\omega_{k,3} = \ek + V_{3}$, with $\ek = - 2J \cos(k)$. The $\delta$-function ensures that excitations are created resonantly: the photon energy of the rf field, $\hbar\omega_{\text{rf}}$, has to match the energy difference between ground state $\ket{GS}$ and excited state $\ket{n}$. Additionally, a transition can only occur if there is a finite matrix element of the perturbing operator $\cop_{k,2}$ between the initial and final states. 

The required energy for an excitation is hence comprised of two parts: the energy of a free particle in the upper band and the energy of an excitation in the lower band, created by the removal of a fermion of species $\ket{2}$.

We thus see that within linear response theory a linear rise of the expectation values is expected and that the slopes are related to the single-particle spectral function. This implies that within the validity regime of the linear response theory,  the rf spectroscopy technique can be used to probe the single-particle spectral functions~\cite{ZwierleinKetterle2008}. Correspondingly, the total transfer rate will be the sum over all momenta of Eq.~\ref{eq:linear_response}, $\dot{N}_{3}(t) \sim \Omega_{23}^{2}\sum_{k} A(k,\omega_{\text{rf}})$.

Information about the spectral functions and in particular of their support can be obtained from the Bethe ansatz. The energy difference is related to the elementary excitations in the Bethe ansatz by $\hbar\omega_{\text{rf}} = \hbar(\omega_{n} - \omega_{0}) + \hbar\omega_{k,3} = \epsilon_{sc}(k) + \hbar\omega_{k,3}$.

Note that the regime of the Rabi oscillations, Eq.\ref{eq:Rabioscillation}, \emph{cannot} be described within linear response calculations. Fundamentally, it is the coupling to a continuous band of levels with different frequencies, which makes Rabi oscillations give way to the linear response regime. We expect stronger interactions to increase the level mixing and thus, to make it easier to reach the linear regime.

\section{Numerically exact solution by the time-dependent matrix product state algorithm}
\label{sec:mps}
In this section, we review the time-dependent matrix product state algorithm which we employ to obtain the quasi-exact solution of the driven interacting many-body problem.

As in experiments, we assume the system to be initially prepared in the ground state of $H_{0}$, Eq.~\ref{eq:H0}. The rf field is applied at $t > 0$. The ground state is obtained using the density matrix renormalization group (DMRG) method in the formulation of matrix product states (MPS). Additionally, the time evolution is performed using the variational time-dependent matrix product (tMPS) approach~\cite{WhiteFeiguin2004,DaleyVidal2004,Schollwoeck2011}. Both approaches are based on using a variational ansatz for the many-body wavefunction of MPS form, i.e.
\begin{equation}
  \ket{\psi[A]} = \sum_{\{\sigma_{i}\}} \tr{(A^{[1]\sigma_{1}}_{1}A^{[2]\sigma_{2}}_{2}\cdots A^{[L]\sigma_{L}}_{L})} \ket{\sigma_{1} \sigma_{2} \cdots \sigma_{L}} \nonumber ~.
\end{equation} 
Any state can be represented in this form with suitable matrix dimension, the so-called bond dimension, of the matrices $A^{[i]\sigma_{i}}_{a_{i-1},a_{i}}$ at chosen $i$ and $\sigma_{i}$. For an exact representation, the local matrices have the bond dimension $D_{i-1}\times D_{i}$, and there are $\sigma_{i} = \{1,\dots,d\}$ of them, where $d$ is the physical dimension of the local Hilbert space. For the present three-species Fermi-Hubbard model, $d=8$. At this point, the above representation is exact, however just like the full Hilbert space of the system grows exponentially with the system size, so too must the bond dimension D of the MPS grow.

In order to generate a numerically feasible treatment, the bond dimensions are cut using a singular value decomposition to a treatable value corresponding to an optimal approximation of the state. 

The bond dimension is directly related to the amount of entanglement between bipartitions of the system along a bond. Fortunately, ground states of one-dimensional, gapped Hamiltonians show an area law entanglement spectrum~\cite{Hastings2004}, which allows for an efficient truncation of the MPS bond dimension and a reduction to polynomial complexity. It is the beneficial entanglement spectrum that allows for the efficient simulation of low-dimensional quantum systems. The discarded weight of singular values (known as the truncation error $\epsilon$), and the bond dimension D together control the accuracy of the MPS simulation. Finally, we approximate the time evolution operator $U(t) = \exp(-iHt/\hbar)$ by a second-order Trotter-Suzuki decomposition~\cite{Trotter1959,Suzuki1976,Suzuki1985,Trotter1991}, controlled by the time step $dt$. Unless stated otherwise, we have chosen $\epsilon = 10^{-12}$, $D = 500$, and $dt = 0.005\hbar/J$ for our simulations, to ensure convergence of our results.

\section{Weakly attractive Hubbard model: response to weak rf driving}
\label{sec:weak_attraction}
In this section, we describe our results on the dynamics induced by a weak rf driving in the weakly attractive Hubbard model. As naively expected, atom transfer from level
$\ket{2}$ to $\ket{3}$ is induced by the rf drive. However, the amplitude and form of this transfer, and the subsequent dynamics depends very much on the rf frequency
and on the interaction strength between the atoms in levels $\ket{1}$ and $\ket{2}$. We analyze in detail the time-evolution of various quantities shedding light on the intricate dynamics of this system. We begin in~\ref{sec:nk3_weakU12} by considering the time dependence of the momentum-resolved transfer to level $|3\rangle$, the quantity illustrating most directly the system dynamics. We discuss in which situations Rabi-like or linear response behaviours are observed. In~\ref{sec:momentum_distribution_weakU12}, we then turn our attention to the analysis of the momentum resolved density distributions for all three levels, and, in~\ref{sec:pair_distribution_weakU12}, present the evolution of pair momentum distribution associated with levels $\ket{1}$ and $\ket{2}$. These quantities provide us with insights into the dynamics of the interacting state induced by the transfer. We conclude this section by commenting, in~\ref{sec:N3_weakU12}, on the evolution and spectrum of the total population transfer to level $\ket{3}$,
which is the experimentally most accessible quantity.

\subsection{Momentum-resolved transfer to the third level}
\label{sec:nk3_weakU12}
\begin{figure} 
  \includegraphics[width=1.0\columnwidth]{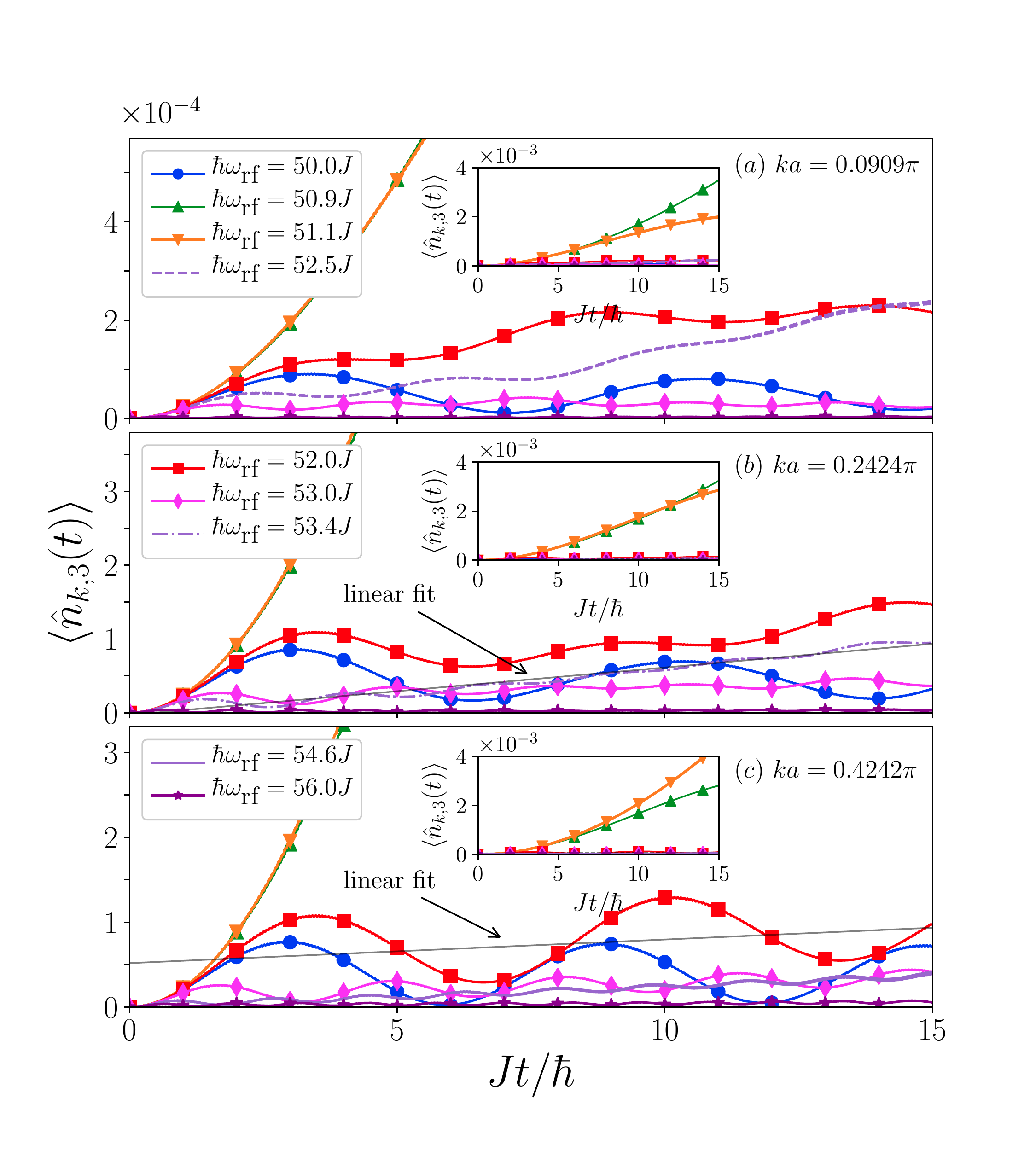}
  \caption{Time-dependence of the upper band population $\expval{\nop_{k,3}(t)}$ for selected momentum states at various driving frequencies $\omega_\text{rf}$ across the
          main resonance for a system of $L = 32$ sites at half-filling for interaction strength $U = -2J$. Level $\ket{3}$ is $V_{3} = 50J$ in energy above levels $\ket{1}$ and $\ket{2}$, and the Rabi frequency is $\hbar \Omega_{23} = 0.01J$. The time-evolution can be separated into two regimes: a Rabi-like regime, occurring in the far red and blue-detuned limits and near the main resonance, and a linear response regime. Full grey lines are examples of linear fits, the extracted slopes are then used to construct the spectrum shown in Fig.~\ref{fig:nk3spectrum_weakU12_bothOmegaR}. (a) $k a = 0.0909\pi$; (b) $k a = 0.2424\pi$; (c) $k a = 0.4242\pi$. 
          Except stated otherwise, curves with the same marker and colour show the same driving frequency in all three panels.
          To ensure convergence of our results, we have separately varied the bond dimension ($D=400$), truncation error ($\epsilon = 10^{-13}$), and time step ($Jdt = 0.002\hbar$) from the parameters given in section~\ref{sec:mps}. The corresponding curves are overlaid for driving frequencies $\hbar\omega_{\text{rf}} = 51.1J$ (a-c), $\hbar\omega_{\text{rf}} = 52.5J$ (a), and $\hbar\omega_{\text{rf}} = 54.6J$ (c). The simulation error is therefore below the linewidth shown.
          }
  \label{fig:nk3_weakU12_weakOmegaR}
\end{figure}

We first analyze the momentum-resolved atom transfer from $|2\rangle$ to $|3\rangle$ with a bare level spacing of $V_{3} = 50J$, induced by the rf driving, by considering the time-evolution of $\expval{\nop_{k,3}(t)}$. We find that the dynamics of $\expval{\nop_{k,3}(t)}$ depends strongly on the momentum, $k$, and the driving frequency, $\omega_\text{rf}$. In Fig.~\ref{fig:nk3_weakU12_weakOmegaR}, we present the time-evolution of this observable for a large window of driving frequencies and three representative momentum values: a value near zero momentum ($k a = 0.0909 \pi$), a value halfway to the Fermi edge ($k a = 0.2424 \pi$) and a value near the Fermi edge ($k a = 0.4242 \pi$). For small frequencies, such as $\hbar \omega_\text{rf} = 50J$ (blue curve in all panels), the transfer is dominated by fast, off-resonant Rabi oscillations with relatively little transfer. $\hbar \omega_{\text{rf}} = V_{3} = 50J$ is the resonance for the non-interacting system, and is now, for $U = -2J$ corresponding to a red-detuned driving with respect to the maximum transfer peak occurring at $\hbar \omega_{\text{rf}} = 51J$. Increasing the driving frequency, the Rabi oscillations become slower and are damped which we attribute to the interaction induced level mixing. At frequencies between $\hbar \omega_\text{rf} = 50.9J$ and $51.1J$ (green and orange curves in all panels), the transfer increases substantially.

In fact, one can see that the largest transfer shifts to higher frequencies with increasing momentum, which we attribute to a resonance. Then, beyond this resonance, the form of the time-evolution changes drastically and is, after an initial slow rise, almost linear (with superposed modulations) over a significant time interval. Increasing the driving frequency even further, the linear behaviour persists, but the slope is in general a decreasing function of the driving frequency. Unexpectedly, a second peak occurs at a second set of driving frequencies whose value depends on the considered momentum. The subtle effect can be made out in Fig.~\ref{fig:nk3_weakU12_weakOmegaR} (a) at $\hbar\omega_\text{rf} = 52.5J$, (b) at $\hbar\omega_\text{rf} = 53.4J$, and (c) at $\hbar\omega_\text{rf} = 54.6J$ (light purple lines without markers). At these frequencies, the slope and transfer surpass previous values at lower driving frequencies. This signals that a second resonance occurs in the transfer. Finally, once the system is driven very far on the blue-detuned side, one recovers again a Rabi-dominated, fast oscillating signal with low net transfer.

\begin{figure} 
  \includegraphics[width=1.0\columnwidth]{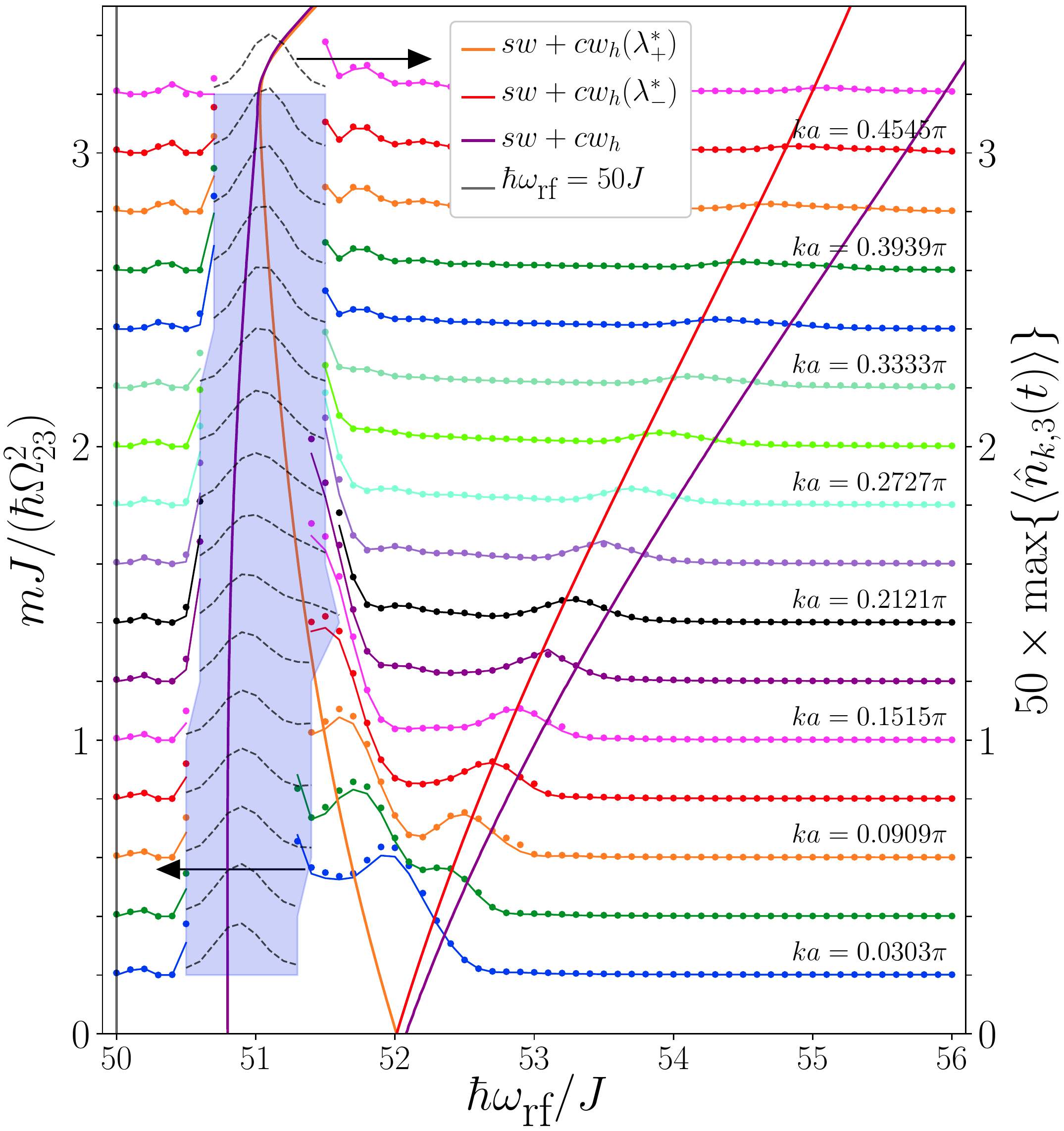}
  \caption{Rescaled momentum-resolved transfer rate to level $|3\rangle$, $m(k,\omega_{\text{rf}})/\Omega_{23}^{2}$, for a system of $L = 32$ sites at half-filling for 
          interaction strength $U = -2J$. Level $|3\rangle$ is $\epsilon_3 = 50J$ in energy above levels $|1\rangle$ and $|2\rangle$. The dots represent the rescaled slopes for $\hbar \Omega_{23} = 0.01J$ and the lines, $\hbar\Omega_{23} = 0.1J$ (left axis, c.f. left arrow). These two data sets are found to be in good agreement. The shaded region corresponds to the frequency interval over which the time-evolution is not linear and the fitting procedure is not possible. In this region, we report the maximum atom transfer in the time interval $0 \leq J t \leq 15\hbar$ (right axis, c.f. right arrow). The momentum values $ka = \frac{m\pi}{L+1}$ are equally spaced and are shifted vertically by $ka (L+1)/(2\pi) = m/2$, where m takes integer values. The bold solid lines are the lower and upper limits of the spin-charge continuum (purple) and two `spin-wave' excitations (orange and red lines) obtained from Bethe ansatz.
          }
  \label{fig:nk3spectrum_weakU12_bothOmegaR}
\end{figure}

This survey of the time-dependence of the momentum-resolved atom transfer clearly shows, for a large window of driving frequencies, that $\expval{\nop_{k,3}(t)}$ rises on average linearly over a fairly long time interval. We can therefore fit $\expval{\nop_{k,3}(t)}$ over this interval as

\begin{align}
  \expval{\nop_{k,3}(t)} &= m(k, \omega_\text{rf})~t + c(k, \omega_\text{rf}), \nonumber
\end{align}
where $m(k, \omega_\text{rf})$ is the slope and $c(k, \omega_\text{rf})$ the intercept. Examples of these fits are displayed in Fig.~\ref{fig:nk3_weakU12_weakOmegaR}. We then report in Fig.~\ref{fig:nk3spectrum_weakU12_bothOmegaR}, as a function of $\omega_\text{rf}$, the rescaled slopes for various momenta and two Rabi frequencies, $\hbar \Omega_{23} = 0.1 J$ and $0.01J$. However, as hinted earlier, close to the strongest resonance, we cannot identify a linear regime, and instead of reporting a slope value, we plot the maximum transfer value to level $\ket{3}$ recorded in the time interval $0 \leq J t \leq 15\hbar$. This region is denoted with shading in Fig.~\ref{fig:nk3spectrum_weakU12_bothOmegaR}. The obtained figure presents well defined features. Even though we cannot identify a linear response regime close to the lower resonance, most of the features can be understood by remembering, from section~\ref{sec:linear_response}, that linear response predicts that the rescaled slope, $m(k, \omega_\text{rf})/\Omega_{23}^2$, is proportional to the single-particle spectral function, $A(k,\omega_\text{rf})$ (see Eq.\ref{eq:linear_response}). Fig.~\ref{fig:nk3spectrum_weakU12_bothOmegaR} can therefore be loosely interpreted as the single-particle spectral function for the attractively interacting Hubbard model. This realization can be put on firmer grounds by overlaying excitation lines predicted from Bethe ansatz taking the upper level dispersion into account such that $\hbar\omega_\text{rf} = \epsilon_{sc}(k) + \hbar\omega_{k,3}$. Here $\epsilon_{sc}(k)$ is the energy of a given excitation inside the spin-charge continuum above the ground state. The purple lines mark the lower and upper edge of the spin-charge continuum (lower-right panel of Fig.~\ref{fig:BetheExcitations_TwoParticle_weakU}). While the other two lines correspond to excitations within the continuum that the drive couples strongly to. These are either mostly of spin-wave (orange and red) character, together with a gapless excitation of the opposite sector, i.e. $\epsilon_{\text{sw}}(q) + \epsilon_{\text{cw}}(\lambda^{*}_{\pm})$, where $p^{\text{h}}_{\text{cw}}(\lambda^{*}_{\pm}) = \pm \frac{\pi}{2a}$.

Comparing further in Fig.~\ref{fig:nk3spectrum_weakU12_bothOmegaR} our numerical results with the Bethe ansatz solutions, we first notice that, for driving frequencies below the spin-charge continuum, the rf photon effectively mainly sees the lower edge of the excitation band (i.e. the bottom of the level $\ket{3}$ band), and the evolution is characterized by off-resonant Rabi oscillations. We also observe that the frequencies marking the onset of the rapid rise of the maximum atom transfer is in very good agreement with the lower edge of the spin-charge continuum. Then, when the drive lies well within the continuum, the transfer rate is finite. As there the drive couples to a continuous band of excitations, levels mix sufficiently, and a linear net transfer emerges. The transfer rate is thus very sensitive to some excitations making up the continuum, as such we observe a pronounced peak when following the spin-wave character lines $\epsilon_{\text{sw}}(q) + \epsilon_{\text{cw}}(\lambda^{*}_{\pm})$ (red and orange lines in Fig.~\ref{fig:nk3spectrum_weakU12_bothOmegaR}). For driving frequencies above the upper edge of the continuum, the slope is reduced and the response goes back to a fast oscillating, low amplitude response, reminiscent of far blue-detuned Rabi oscillations. In this case, the energy conservation condition of Eq.~\ref{eq:linear_response} cannot be strictly fulfilled in this two-particle excitation sector, the rf photon provides too much energy to resonantly excite spin and charge degrees of freedom leading to insufficient coupling and very weak net transfer rates. With yet higher energies of the rf photon we would expect to eventually enter and resonantly couple to the $2n$-particle ($n > 1$) excitation sectors.

\begin{figure} 
  \includegraphics[width=1.0\columnwidth]{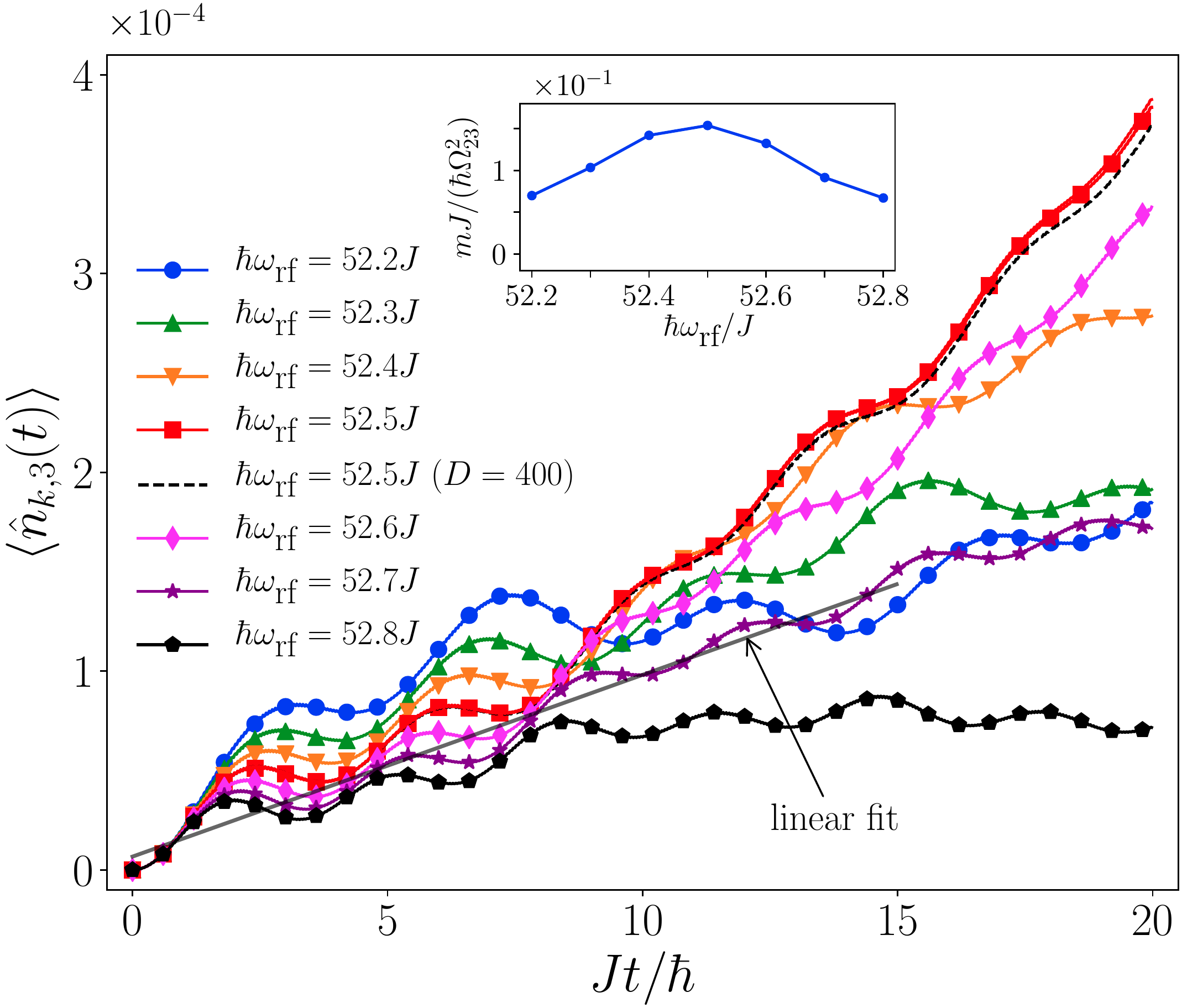}
  \caption{Time-dependence of $\expval{\nop_{k,3}(t)}$ at $k a = 0.0909\pi$ near the resonance at $\hbar \omega_\text{rf} = 52.5J$ for a half-filled system of size $L = 32$ 
          with interaction strength $U = -2J$ and $\hbar\Omega_{23} = 0.01J$. The system coupling to a `spin-wave' excitation translates into an increase of the transfer rate near the resonance and in a fanning out of the curves around  $J t \sim 8 \hbar$. The full grey line is an example of linear fit, the extracted slope is reported in the inset showing a momentum slice of the single-particle excitation spectrum. To ensure convergence of our results, we have separately varied the bond dimension ($D=400$), truncation error ($\epsilon = 10^{-13}$), and time step ($Jdt = 0.002\hbar$) from the parameters given in section~\ref{sec:mps}. The corresponding curves are overlaid for a driving frequency of $\hbar\omega_{\text{rf}} = 52.5J$. The convergence for the time step and truncation error are plotted in the same colour (and the respective error is below the linewidth), while the bond dimension is explicitly shown as a black dashed line.
          }
  \label{fig:nk3_weakU12_weakOmegaR_cluster_L32}
\end{figure}

Due to the strength of the main resonance occurring around $\hbar \omega_\text{rf} = 51.1J$, the other resonances highlighted by red and orange lines in Fig.~\ref{fig:nk3spectrum_weakU12_bothOmegaR} can be more difficult to identify. To remedy this, we present in Fig.~\ref{fig:nk3_weakU12_weakOmegaR_cluster_L32} the evolution of $\expval{\nop_{k,3}(t)}$ for one particular momentum, $k a = 0.0909\pi$, around the `spin-wave' excitation peaking, in this case, at $\hbar \omega_\text{rf} = 52.5J$. As the rf driving couples strongly to this excitation, we see that the character of the dynamics changes as one approaches this resonance. Sufficiently far on both sides of $\hbar \omega_\text{rf} = 52.5J$, the curves collapse onto each other and the transfer is very similar from one driving frequency to the next. As the driving frequency gets closer to the resonance, the transfer rate noticeably increases: one sees in Fig.~\ref{fig:nk3_weakU12_weakOmegaR_cluster_L32} that the curves fan out around $J t = 5 \hbar$ as the ones near the resonance have steeper slopes. We observe this behaviour across all momenta, when following the two excitation lines.

\subsubsection{Finite size effects: $L = 20$}
\begin{figure} 
  \includegraphics[width=1.0\columnwidth]{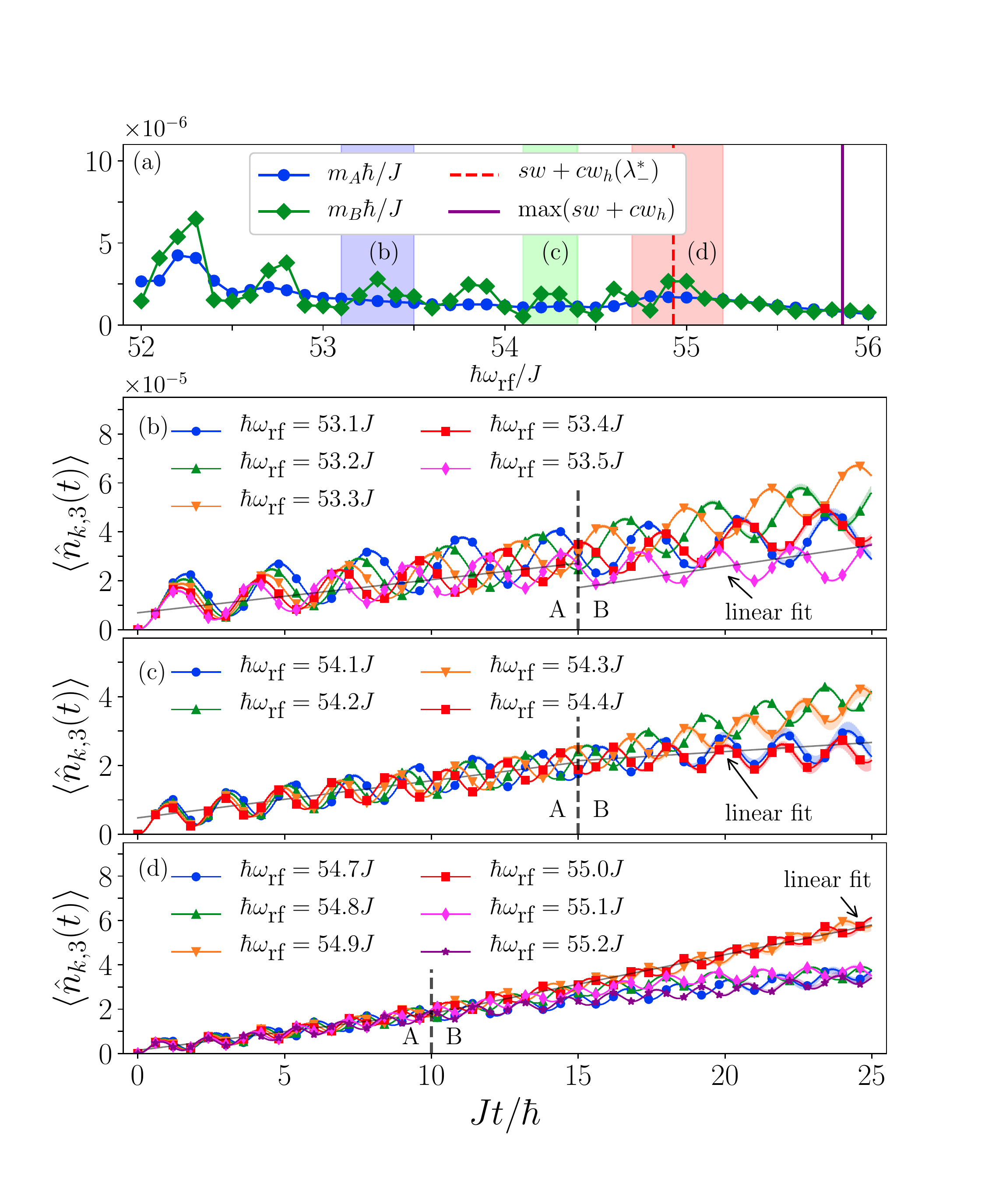}
  \caption{(a) Rescaled momentum-dependent transfer rates to level $|3\rangle$, $m_{A}$ and $m_{B}$, for a system of size $L=20$, $\hbar\Omega_{23} = 0.01J$ and $U = -2.0J$
          for $k a = 0.4762 \pi$. $\expval{\nop_{k,3}(t)}$ for the driving frequencies corresponding to the shaded regions (b), (c) and (d) is shown in the corresponding panels. In (a) the red dashed vertical line marks the position of the `spin-wave' excitation and the purple dashed vertical line marks the upper limit of the spin-charge continuum. In (c), (b) and (d), the solid grey lines are examples of linear fits while the vertical dashed lines mark the boundary between the two fitting regions ``A'' and ``B''. 
          To ensure convergence of our results in (b-d), we have separately varied the bond dimension ($D=600$), truncation error ($\epsilon = 10^{-13}$), and time step 
          ($Jdt = 0.002\hbar$) from the parameters given in section~\ref{sec:mps}. The maximal and minimal deviation is plotted as a shaded region around each curve. Where such a region is not discernible, the numerical error is below the linewidth shown.
          }
  \label{fig:weakU12_weakOmegaR_nk3spectrum_and_cluster_L20}
\end{figure}

Our results for a half-filled system for system sizes of $L=32$ and $L=20$ are in very good agreement. The spectra and excitation peaks we detect show the same features. However for the smaller system, we observe further peaks in the spectra. While for a half-filled $L = 32$ system, this fanning out occurs only along two well defined excitation lines, for smaller systems the situation is different. For $L = 20$ systems, we find several occurrences of this behaviour within the spin-charge continuum. This situation is illustrated in Fig.~\ref{fig:weakU12_weakOmegaR_nk3spectrum_and_cluster_L20} for $k a = 0.4762\pi$. For this momentum, one sees that the time-evolution of $\expval{\nop_{k,3}(t)}$ is split into two regimes: for early times (a time interval denoted as ``A'') the different curves overlap, whereas for later times (a region denoted as ``B'') the curves begin to fan out. In Fig.~\ref{fig:weakU12_weakOmegaR_nk3spectrum_and_cluster_L20} (a), we report the slopes, $m_{A}$ and $m_{B}$, for both regions. When considering $m_{B}$, we find that the spectrum presents oscillations throughout the spin-charge continuum. While the peak at $\hbar \omega_\text{rf} = 54.9J$ is expected as a `spin-wave' excitation occurs at this energy, the reason behind the existence of the other peaks around, for example, $\hbar \omega_\text{rf} = 53.3J$ and $\hbar \omega_\text{rf} = 54.2J$, is not as obvious. One further notices that, for these two peaks, the fanning out occurs at a later time compared to panel (d). In fact, when comparing to the evolution for the system of size $L = 32$, we notice that, for the `spin-wave' excitation the time at which the fanning out occurs has only slightly decreased for the larger system size while similar structures are totally absent at other driving frequencies within the spin-charge continuum. Such fanning out would probably take place at times larger than $Jt = 25 \hbar$. We therefore associate these oscillations to finite size effects, similar behaviours were observed in~\cite{MasselTorma2009} where the timescale marking the beginning of the fanning out was shown to be related to the inverse finite-size gap.

\subsection{Evolution of the momentum distributions}
\label{sec:momentum_distribution_weakU12}
To gain further insight into the way the rf drive is exciting the system, we turn to the evolution of the momentum distributions. As explained previously, atoms in level $\ket{1}$ are not directly coupled by the rf drive, and so their dynamics is entirely induced by the interaction with $\ket{2}$. Furthermore, within linear response we expect to see no change in the momentum distribution $\nop_{k,1}$, thus all changes we observe in $\ket{1}$ are effects \emph{beyond} linear response.

Whilst the transfer and thus the changes in $\expval{\nop_{k,2}}$ and $\expval{\nop_{k,3}}$ depend strongly on the momentum and rf frequency, the changes in level $\ket{1}$ are mostly around the Fermi edge. For weak Rabi coupling, $\hbar\Omega_{23}=0.01J$, the net transfer is very small, so the absolute momentum density distribution $\expval{\nop_{k,\sigma}}$, with $\sigma = \{1,2\}$, is only slightly altered during the evolution. For $\sigma = \{1,2\}$, the initial distributions have a step-like profile smoothed out by the effect of interaction, and a $\sim 80\%$ drop in occupation around the Fermi edge. Looking therefore at the \emph{deviations} of the momentum distribution $\expval{\nop_{k,\sigma}(t) - \nop_{k,\sigma}(0)}$, reveals the detailed effect of the dynamics. We will now look at the different cases of driving frequencies below (red-detuned), above (blue-detuned), and on resonance ($\hbar \omega_{\text{rf}}=51.1J$) to explain this structure in detail.

\begin{figure} 
  \includegraphics[width=1.0\columnwidth]{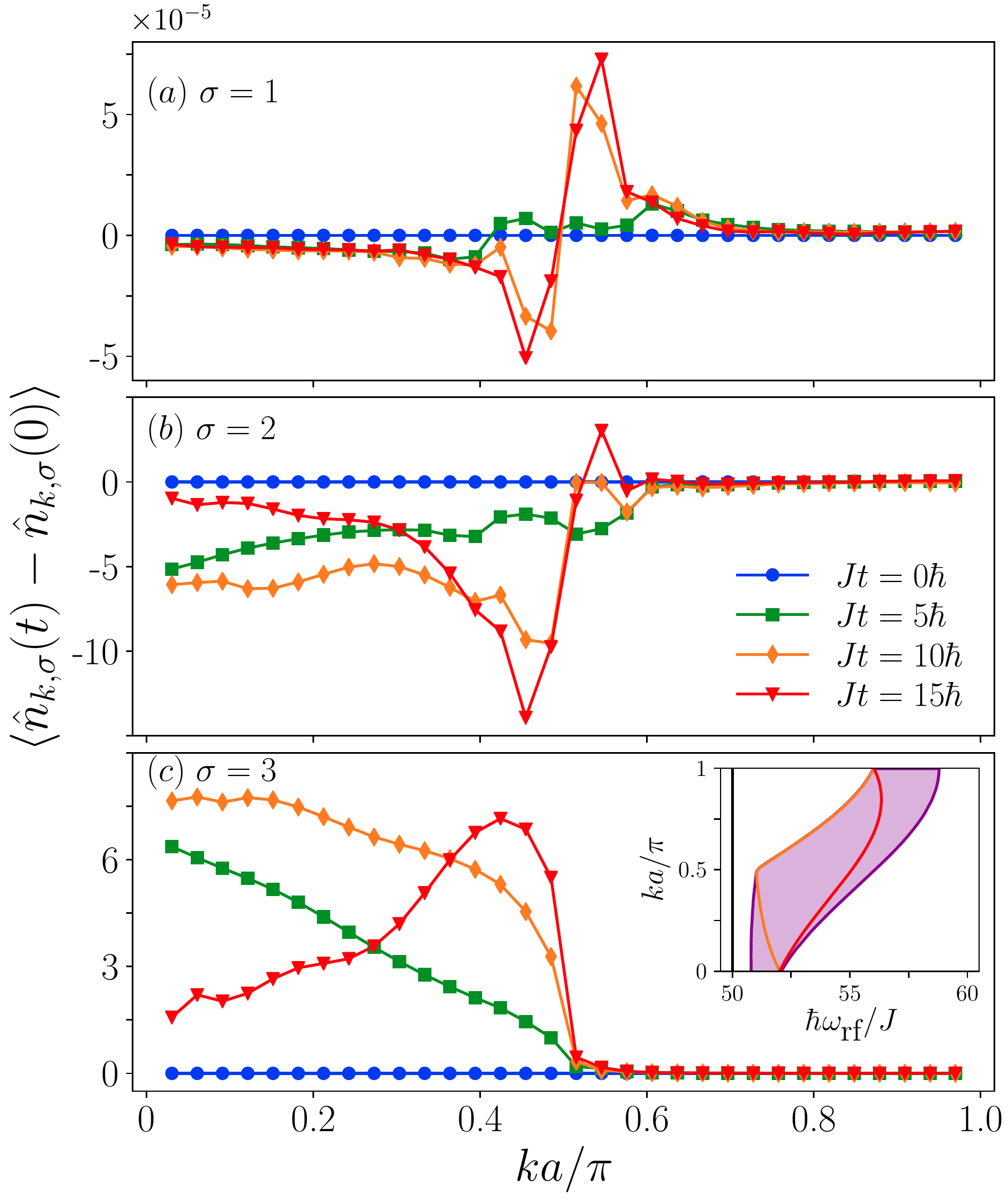}
  \caption{Evolution of the differences of the momentum distributions at times $t$ compared to the initial time, $\expval{\nop_{k,\sigma}(t)}-\expval{\nop_{k,\sigma}(0)}$, 
          for $U = -2J$, $\hbar \Omega_{23} = 0.01J$, and $\hbar \omega_{\text{rf}} = 50.0J$ for $\sigma=\{1,2,3\}$ in panels (a-c) respectively. We show the different times as marked in the legend in (b). The inset in (c) shows the spin-charge excitation continuum (purple region bounded by purple lines), along with two particular excitations of `spin-wave' character (orange and red lines). The black vertical line marks the driving frequency $\omega_{\textrm{rf}}$.
          }
    \label{fig:FermiEdge_weakU12_weakOmegaR_red}
\end{figure}

Fig.~\ref{fig:FermiEdge_weakU12_weakOmegaR_red} shows the momentum distribution for a red-detuned drive $\hbar\omega_{\textrm{rf}}=50J$. $\ket{1}$ is only significantly affected close to the Fermi edge, where a redistribution of particles from below to above the Fermi step, accompanied by oscillations, takes place, as can be seen in (a). In comparison, (b) shows the response of $\sigma=2$, and we observe that all momenta below the Fermi edge are depleted. Finally, (c) shows the distribution for $\ket{3}$, as shown in real-time in Fig.~\ref{fig:nk3_weakU12_weakOmegaR}. The initial transfer is larger (with faster Rabi oscillations) for smaller momenta. This can be explained by looking at the excitation spectrum as shown in the inset. The driving frequency is red-detuned from all excitations, with an effective, momentum dependent detuning from the lower continuum edge (shaded region). The lower edge of the continuum has a small curvature to higher energies, thereby effectively increasing the detuning with momentum, leading to faster oscillations with lower amplitude for momentum states towards the Fermi edge, consistent with the intuition gained when considering the driving of a non-interacting system in section~\ref{sec:noninteracting_system}.

\begin{figure}
  \includegraphics[width=1.0\columnwidth]{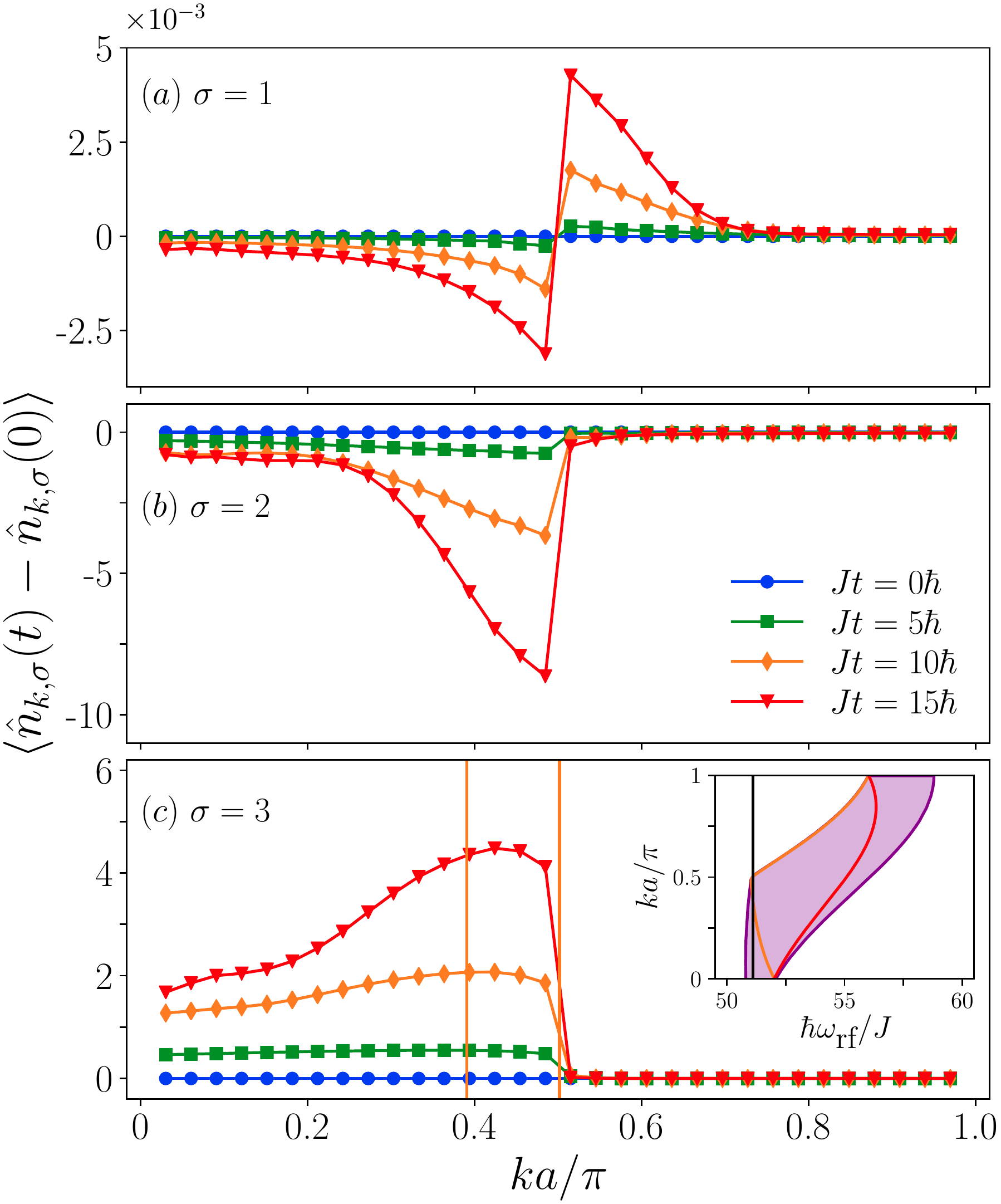}
  \caption{Evolution of the differences of the momentum distributions at times $t$ compared to the initial time, $\expval{\nop_{k,\sigma}(t)}-\expval{\nop_{k,\sigma}(0)}$, 
          for $U = -2J$, $\hbar \Omega_{23} = 0.01J$, and $\hbar \omega_{\text{rf}} = 51.1J$ for $\sigma=\{1,2,3\}$ in panels (a-c) respectively. We show the different times as marked in the legend in (b). The inset in (c) shows the  spin-charge excitation continuum (purple region bounded by purple lines), along with two particular excitations of `spin-wave' character (orange and red lines). The black vertical line marks the driving frequency $\omega_{\textrm{rf}}$ (inset), while the orange vertical lines in panel (c) mark the momenta, at which the driving frequency $\omega_{\textrm{rf}}$ is resonant with the `spin-wave' excitation (orange line, inset). 
          To ensure convergence of our results, we have separately varied the bond dimension ($D=400$), truncation error ($\epsilon = 10^{-13}$), and time step 
          ($Jdt = 0.002\hbar$) from the parameters given in section~\ref{sec:mps}. The maximal and minimal deviation is shown as a shaded region around the corresponding curve (same colour respectively), and if not discernible, the numerical error is below the linewidth.
          }
  \label{fig:FermiEdge_weakU12_weakOmegaR_res}
\end{figure}

\begin{figure} 
  \includegraphics[width=1.0\columnwidth]{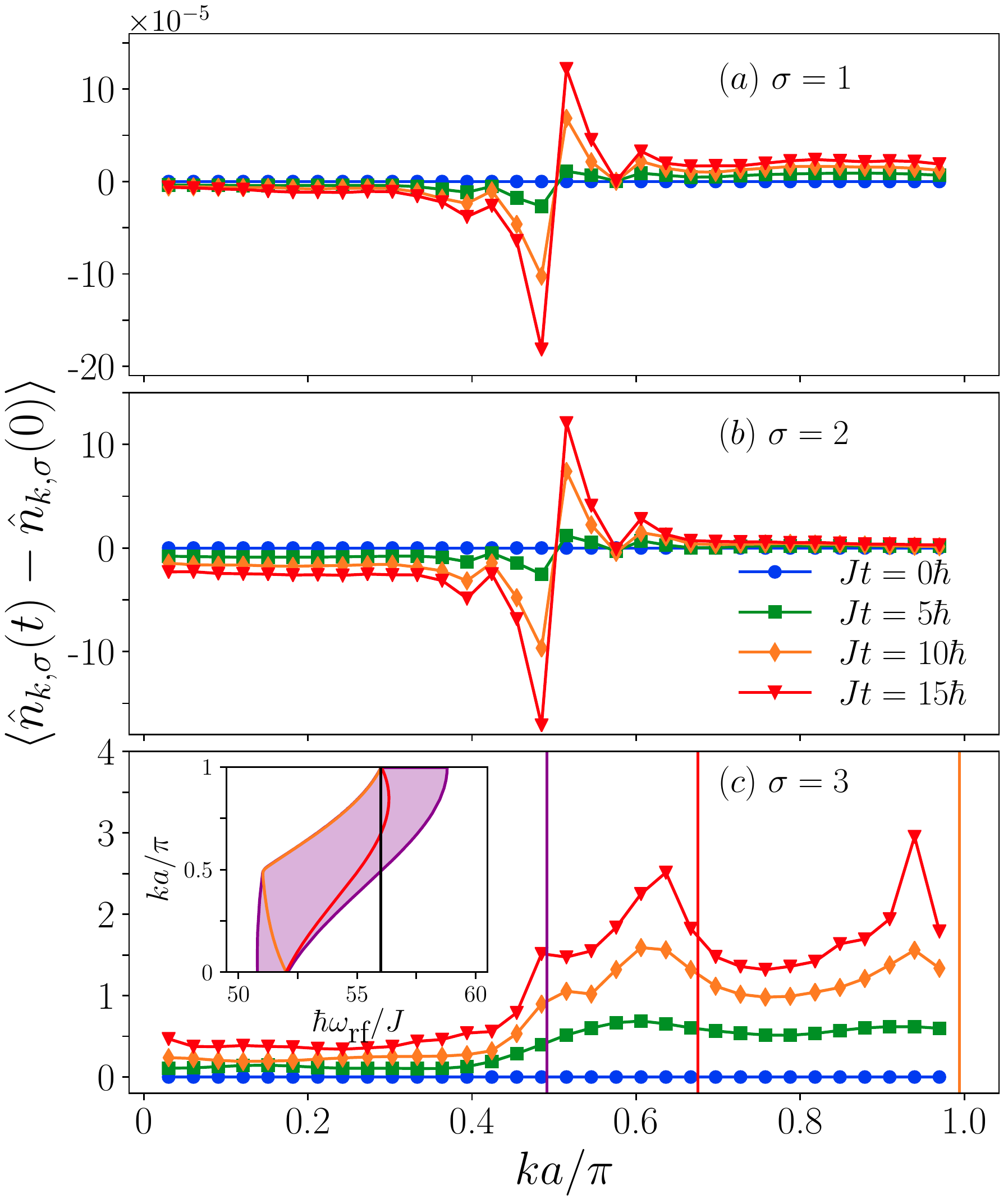}
  \caption{Evolution of the differences of the momentum distributions at times $t$ compared to the initial time, $\expval{\nop_{k,\sigma}(t)}-\expval{\nop_{k,\sigma}(0)}$,
          for $U = -2J$, $\hbar \Omega_{23} = 0.01J$, and $\hbar \omega_{\text{rf}} = 56.0J$ for $\sigma=\{1,2,3\}$ in panels (a-c) respectively. We show the different times as marked in legend in (b). The inset in (c) shows the spin-charge excitation continuum (purple region bounded by purple lines), along with two particular excitations of `spin-wave' character (orange and red lines). The black vertical line marks the driving frequency $\omega_{\textrm{rf}}$ (inset), while the coloured vertical lines in panel (c) mark the momenta, at which the driving frequency $\omega_{\textrm{rf}}$ is resonant with either the `spin-wave' excitations (red and orange), or the upper spin-charge continuum edge (purple).
          }
  \label{fig:FermiEdge_weakU12_weakOmegaR_blue}
\end{figure}

For driving frequencies near the main resonance as shown in Fig.~\ref{fig:FermiEdge_weakU12_weakOmegaR_res}, $\expval{\nop_{k,2}}$ (b) is depleted asymmetrically just below the Fermi edge, while $\expval{\nop_{k,1}}$ (a), purely an interaction effect, shows an almost symmetric response around the same momentum value. The Rabi oscillations in $\expval{\nop_{k,3}}$ (c) appear to be largely in phase, but with an amplitude that is increasing towards larger momenta. Again referring to the inset in (c), the curvature of the lower continuum edge means, that whilst we are driving $k a \sim \pi/2$ almost resonantly, the drive is already slightly above the $k \sim 0$ resonance, which leads to a reduced, but finite transfer rate for all momenta below the Fermi edge. For momenta above the Fermi edge as for the $\sigma=1$ level, the occupation mainly stems from a redistribution induced by the interaction within level $\sigma=1,2$. Compared to off-resonants drives as in Fig.~\ref{fig:FermiEdge_weakU12_weakOmegaR_red} or Fig.~\ref{fig:FermiEdge_weakU12_weakOmegaR_blue}, the transfer is greatly enhanced by up to two orders of magnitude, and the (a)symmetric depletion of $\ket{2}$ ($\ket{1}$) is very strongly pronounced and clearly visible.

\begin{figure}
  \includegraphics[width=1.0\columnwidth]{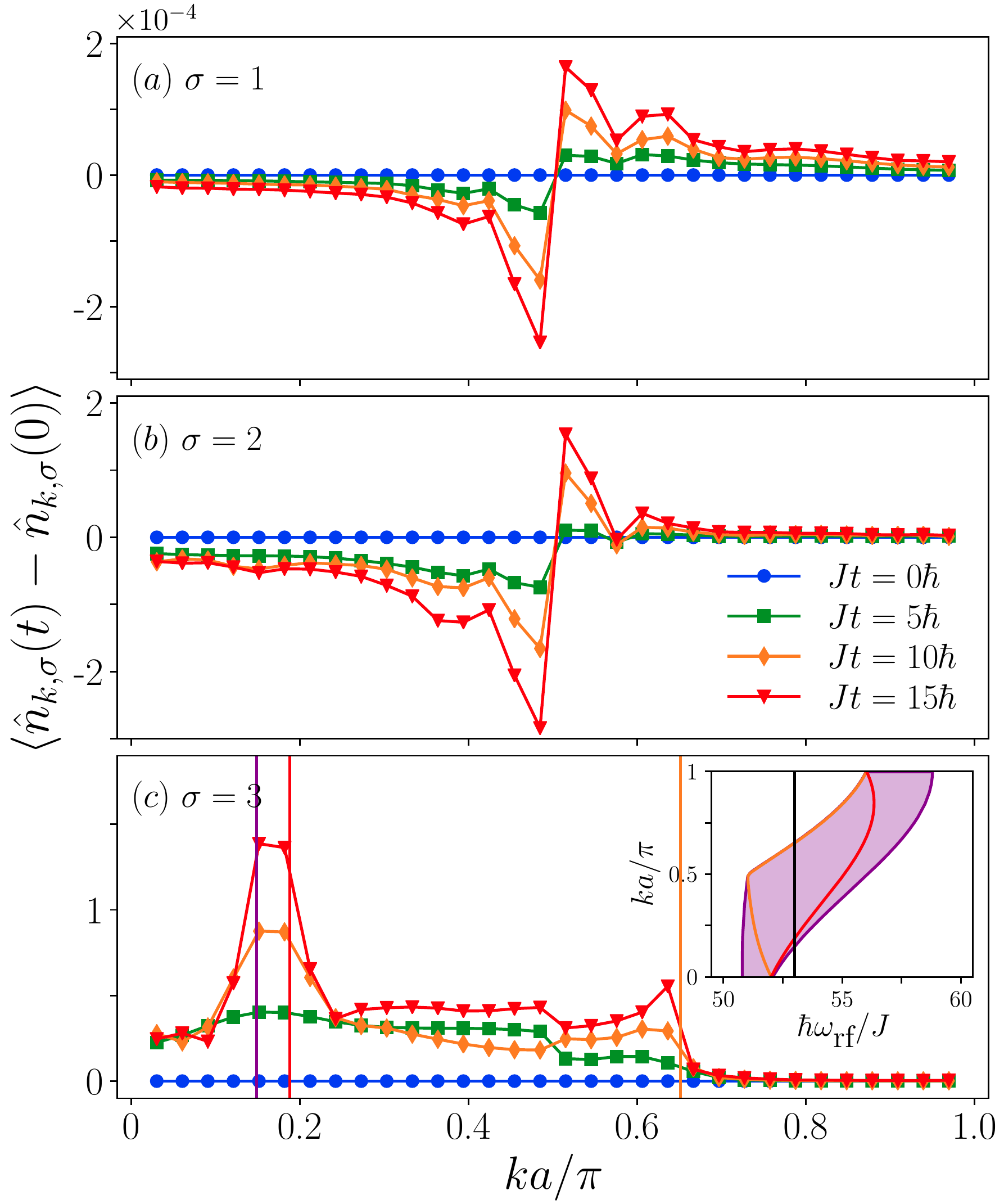}
  \caption{Evolution of the differences of the momentum distributions at times $t$ compared to the initial time, $\expval{\nop_{k,\sigma}(t)}-\expval{\nop_{k,\sigma}(0)}$,
          for $U = -2J$, $\hbar \Omega_{23} = 0.01J$, and $\hbar \omega_{\text{rf}} = 53.0J$ for $\sigma=\{1,2,3\}$ in panels (a-c) respectively. We show the different times as marked in the legend in (b). The inset in (c) shows the spin-charge excitation continuum (purple region bounded by purple lines), along with two particular excitations of `spin-wave' character (orange and red lines). The black vertical line marks the driving frequency $\omega_{\textrm{rf}}$ (inset), while the coloured vertical lines in panel (c) mark the momenta, at which the driving frequency $\omega_{\textrm{rf}}$ is resonant with either the `spin-wave' excitations (red and orange), or the upper spin-charge continuum edge (purple).
          }
  \label{fig:FermiEdge_weakU12_weakOmegaR_inside}
\end{figure}

For Fig.~\ref{fig:FermiEdge_weakU12_weakOmegaR_inside} and Fig.~\ref{fig:FermiEdge_weakU12_weakOmegaR_blue}, the driving frequency, $\omega_\text{rf}$, lies well within the continuum. Interestingly, $\expval{\nop_{k,3}}$ (c) reveals a very distinct response not seen before. The momentum distribution develops a strong two-peak structure at $ka \sim \{0.18\pi,0.63\pi\}$. From the inset in (c) we see that these peaks corresponds to the excitation of particular excitation lines of `spin-wave' character, $\epsilon_{\text{sw}} + \epsilon_{\text{cw}}(\lambda^{*}_{-})$ (i.e. a spin-wave with a gapless charge-wave), at two distinct momenta resonantly. We can thereby identify the peaks developed in $\expval{\nop_{k,3}}$ as a signature of the activation of these excitations in the system. The observed peaks are monotonically growing in time. With increasing $\omega_{\textrm{rf}}$ the two-peak structure shifts to larger momenta, so that in Fig.~\ref{fig:FermiEdge_weakU12_weakOmegaR_blue} they are both above the initial Fermi edge. This explains the very small amplitudes observed for the changes.  We note that, while the Bethe ansatz does not allow easy analytical access to matrix elements, monitoring the dynamics of the density momentum distribution provides important information about the underlying structure of the corresponding matrix elements, and furthermore exhibits dynamical effects beyond linear response calculations as seen in $\expval{\nop_{k,1}}$.

The upper two panels (a-b) of Fig.~\ref{fig:FermiEdge_weakU12_weakOmegaR_inside} and Fig.~\ref{fig:FermiEdge_weakU12_weakOmegaR_blue}, show the population redistribution of $\expval{\nop_{k,\sigma=1,2}}$. The density redistribution has two effects. The dominant one stems from the physical transfer between levels $\ket{2}$ and $\ket{3}$, and can be clearly seen in panel (b) for $\expval{\nop_{k,2}(t)}$ as the occupation decreases below the Fermi edge. The redistribution of populations due to the interaction and scattering between atoms is the only channel that affects the density distribution $\expval{\nop_{k,1}}$. For these two blue-detuned drivings, the occupation in $\expval{\nop_{k,1}}$ reveals signatures of the resonant coupling to excitations of `spin-wave' character. In Fig.~\ref{fig:FermiEdge_weakU12_weakOmegaR_inside} (a), one sees that a secondary peak is developing near the momentum value corresponding to the crossing of the upper (orange) excitation line. This driving noticeably perturbs the system beyond a simple occupation redistribution around the Fermi edge. The situation is similar for the driving at $\hbar \omega_\text{rf} = 56J$ shown in Fig.~\ref{fig:FermiEdge_weakU12_weakOmegaR_blue}. In this case, a secondary peak develops at a momentum approximately corresponding to the crossing of the lower (red) excitation line. Revealing this secondary peak at larger momentum values requires a redistribution that would likely not be captured within linear response.

\subsection{Evolution of the pair distribution}
\label{sec:pair_distribution_weakU12}
As already discussed, the ground state of the Fermi-Hubbard model for attractive interactions presents superconducting correlations. Therefore, in this section we briefly comment on how the rf drive influences the superconducting pairing. We find that, in contrast to the momentum distribution discussed above, mostly small momenta $k \sim 0$ of the pair distribution are affected and changed by the rf drive. In order to show this, we investigate the evolution of the superconducting pair correlations by analyzing the pair structure factor given by

\begin{equation}
  P_{k}(t) = \frac{1}{L} \sum_{i,j} e^{ik(r_{i} - r_{j})} \bra{\Psi(t)}
  \left( \hat{\Delta}^{\dagger}_{i} \hat{\Delta}^{\phantom{\dagger}}_{j} + \text{h.c.}\right) \ket{\Psi(t)}, \nonumber
  \label{eq:pair_distribution}
\end{equation}
where $\ket{\Psi(t)}$ is the evolved wavefunction, $\hat{\Delta}^{\phantom{\dagger}}_{i} = \cop_{i,1}\cop_{i,2}$, the pair annihilation operator at site $i$, and we use here the exponential Fourier transform, i.e. $k = \frac{2\pi n}{L}$ with discretisation $n = \{-\frac{L}{2}+1, \ldots, \frac{L}{2}\}$, because it mimics the time-of-flight imaging in cold atom experiments.

\begin{figure} 
	\includegraphics[width=1.0\columnwidth]{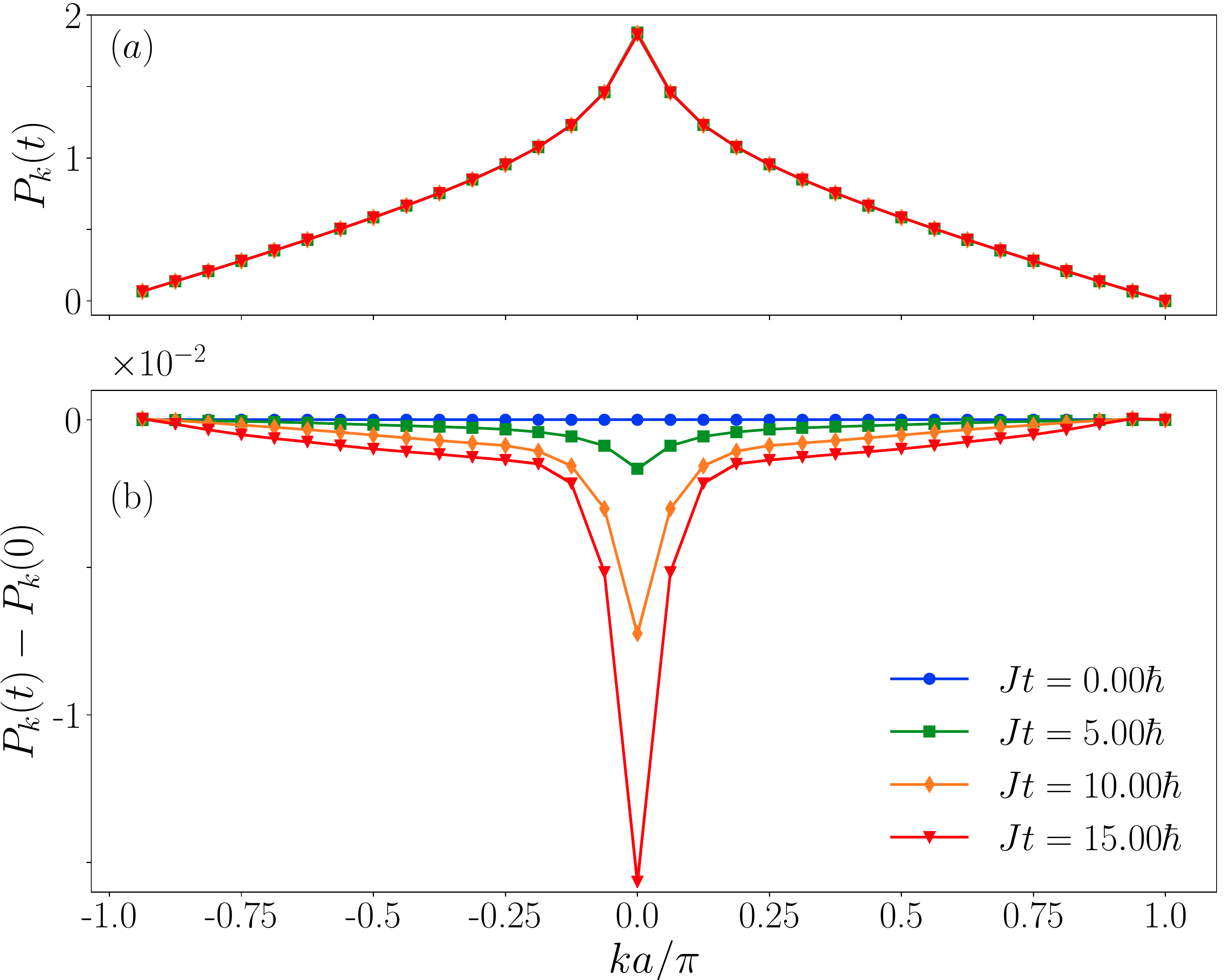}
	\caption{Evolution of (upper panel) the pair distribution $P_{k}(t)$ and (lower panel) the deviation of the pair distribution from its initial value,$P_{k}(t)-P_{k}(0)$, 
          for the attractive Hubbard model of size $L = 32$ at half-filling for $U = -2J$, $\hbar \Omega_{23} = 0.01J$ and $\hbar \omega_\text{rf} = 51.1J$.
          To ensure convergence of our results, we have separately varied the bond dimension ($D=400$), truncation error ($\epsilon = 10^{-13}$), and time step  
          ($Jdt = 0.002\hbar$) from the parameters given in section~\ref{sec:mps}. The corresponding curves are overlaid in the same colours as the shown data. The numerical error is therefore below the linewidth.
          }
	\label{fig:PairDistribution_weakU12_weakOmegaR_res}
\end{figure}

Fig.~\ref{fig:PairDistribution_weakU12_weakOmegaR_res} shows the pair correlator for different points in time for resonant driving. Considering first the absolute pair correlation, $P_k(t)$, one sees in panel (a) that this quantity is only mildly affected by the driving. Hence, due to the weak driving amplitude, we monitor instead the deviation to the initial state (b). We find a weak background depletion for all momenta, which is however overshadowed by the stronger reduction at $k=0$. This is in stark contrast to the involved structure of $\expval{\nop_{k,\sigma}}$ of the previous section. Here, during the evolution, even on resonance, the pair correlation amplitude is monotonically decreased for all momenta (particularly for $|ka|$ away from $\pi$). Similarly to previous observations, the change in the pairing correlations is nearly two orders of magnitude larger on resonance compared to off-resonant driving frequencies.

The rf drive creates superposition of $\ket{2}$ and $\ket{3}$ particles and injects energy into the system. During this process, pairs making up the superconducting state are altered, and excitations, in the manifold formed from levels $\ket{1}$ and $\ket{2}$, are created. Within our model, due to the absence of any dissipation channels, the system cannot relax back into the ground state and can be seen as heating up. Moreover, when the atoms are transferred back from $\ket{3}$ to $\ket{2}$, they are no longer coherent with the $\ket{1}$ atom they originally formed a pair with, the decoherence accumulating with time. We therefore conclude that the rf drive induces decoherence and causes heating, leading to a suppression of the superconducting pair correlations $P_{k}(t)$, as observed in (b).

\subsection{Total transfer to the third state}
\label{sec:N3_weakU12}
Experimentally, the simplest observable to detect is the total transfer to the third state, $N_{3}(t)$. We discuss in this section, which information can already be extracted from this quantity. 

\begin{figure} 
  \includegraphics[width=1.0\columnwidth]{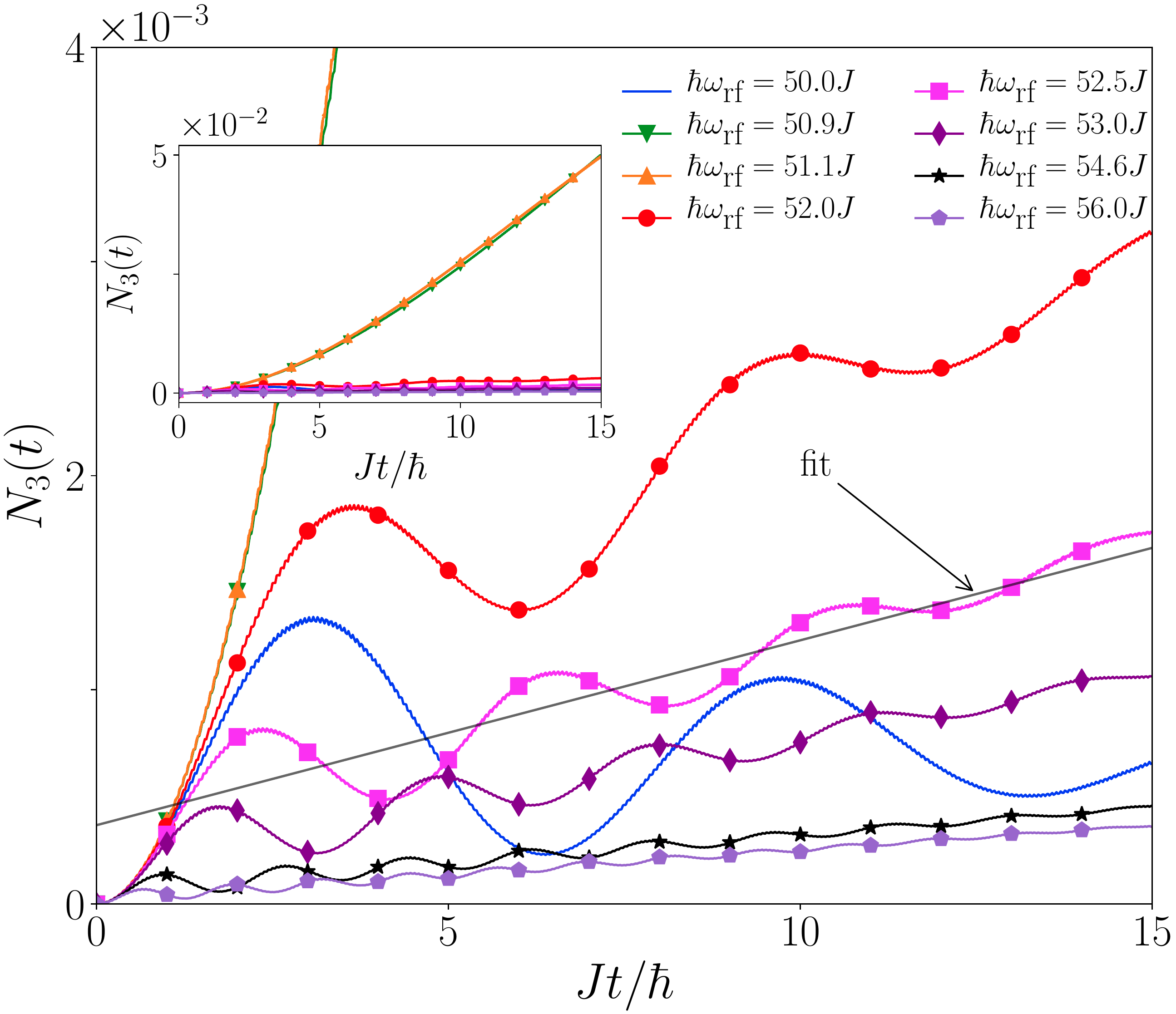}
  \caption{Time-dependence of the total transfer to the third state, $N_{3}(t)$, for the attractive Hubbard model of size $L=32$ at half-filling for $U = -2J$,
          $\hbar\Omega_{23} = 0.01J$ and several driving frequencies $\omega_{\textrm{rf}}$. The main plot focuses on the curves for off-resonance driving, whilst the inset puts these in context when compared to resonantly driven situations. The grey solid line is the linear portion of a fit to the time-dependence for $\hbar\omega_\text{rf} = 52.5J$. 
          To ensure convergence of our results, we have separately varied the bond dimension ($D=400$), truncation error ($\epsilon = 10^{-13}$), and time step ($Jdt = 0.002\hbar$) from the parameters given in section~\ref{sec:mps}. The corresponding curves for driving frequencies $\hbar\omega_{\text{rf}}=51.1J$ and $52.5J$ are overlaid in the same colours as the shown data. The simulation error is therefore below the linewidth.
          }
  \label{fig:N3_weakU12_smallOmegaR}
\end{figure}

The total transfer to the upper level is shown in Fig.~\ref{fig:N3_weakU12_smallOmegaR} for different driving frequencies. Since the total transfer is comprised of the sum of the momentum resolved transfers, we expect to recover the same physics as discussed previously in section~\ref{sec:nk3_weakU12}. Indeed, well below the resonance, for $\hbar\omega_{\textrm{rf}} \lesssim 50.7J$, the total transfer shows Rabi oscillations around a small value. At resonance, $\hbar\omega_{\text{rf}} \sim 51J$, the transfer is maximal (see inset), showing a slow and large amplitude oscillation, while on the blue-detuned side of the resonance the evolution is characterised by a linear steady increase in the total population of the upper level (with oscillations superposed on top). In this situation, we are driving excitations inside the spin-charge continuum and hence coupling to a band of states.

\begin{figure} 
  \includegraphics[width=1.0\columnwidth]{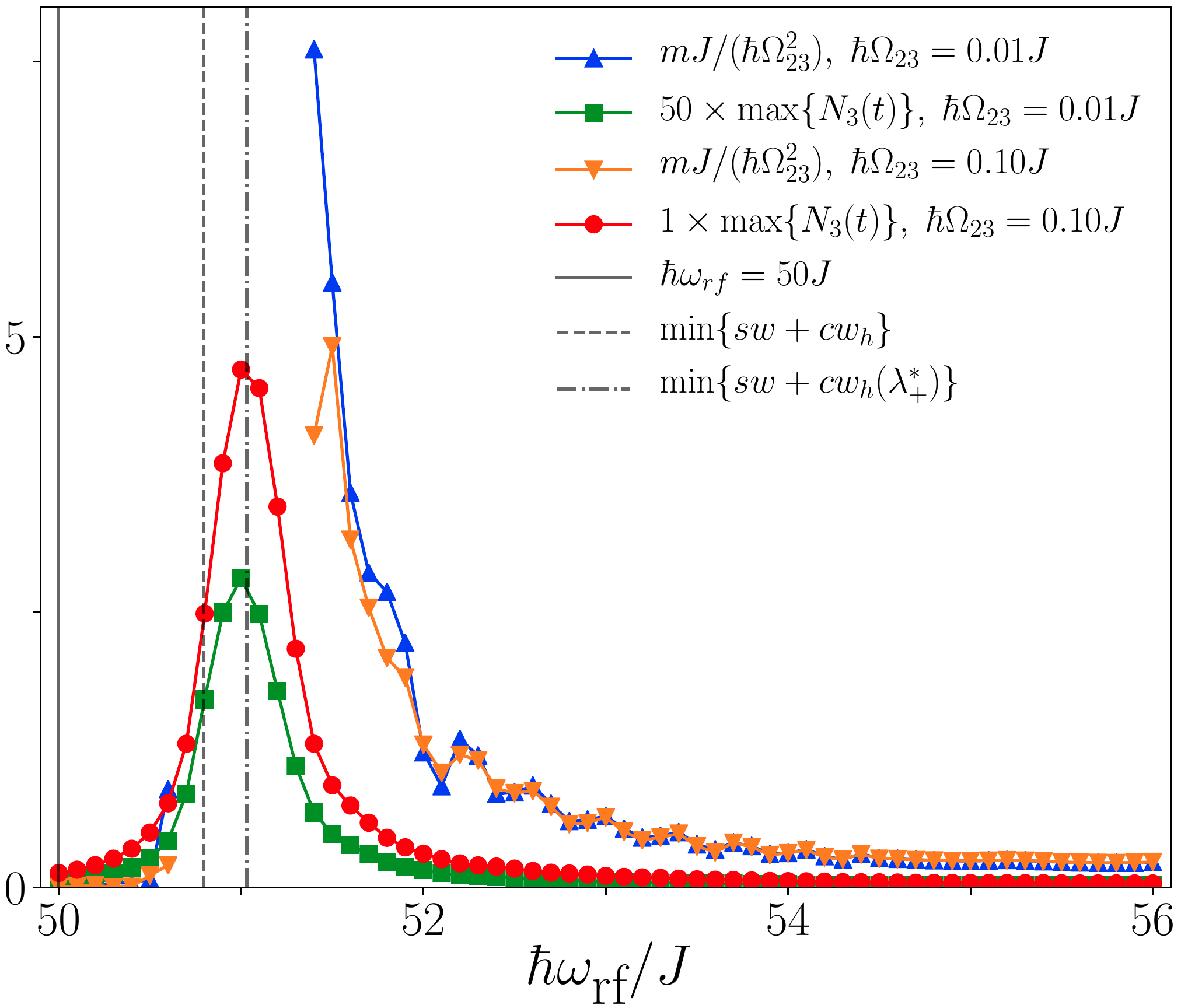}
  \caption{Radio-frequency transfer rate (scaled by the squared Rabi frequency) and scaled maximum net transfer between $0 \leq Jt \leq 15\hbar$ for a system of size $L = 32$,
          $U = -2J$, and  $\hbar\Omega_{23} = 0.01J$ and $0.1J$. The grey solid line indicates the resonance position for a non-interacting system, the grey dashed line marks the lower edge of the spin-charge excitation continuum, whilst the grey dash-dotted line highlights the minimum energy of the `spin-wave'-type excitations, $\epsilon_{sw}(q) + \epsilon_{cw}(\lambda^{*}_{+})$, where $p^{h}_{cw}(\lambda^{*}_{+}) = \pi/(2a)$.
          }
  \label{fig:N3spectrum_weakU12_bothOmegaR}
\end{figure}

Fig.~\ref{fig:N3spectrum_weakU12_bothOmegaR} shows the rescaled slope, $m/\Omega_{23}^{2}$, extracted from fits of the form $N_{3}(t) = m~t + A\cos(\omega t)e^{-\gamma t} + c$, where $m$, $A$, $\omega$, $\gamma$ and $c$ are all fitting parameters, for two different drivings $\hbar\Omega_{23} = 0.01J$ and $0.1J$. Both driving amplitudes exhibit a clear resonance at $\hbar\omega_{\text{rf}} \sim  51J$. For blue detuning of the rf field from this resonance, the two curves collapse onto each other, as already seen in Fig.~\ref{fig:nk3spectrum_weakU12_bothOmegaR}. This indicates that the evolution has entered the linear regime, and confirms the validity of using the linear response approach for these frequencies. Discrepancies between the curves arise close to the resonance, where the transfer is maximal and dominated by slow Rabi oscillations, not captured by the linear response calculations. Our fits do not cover this regime for two reasons. First, the transfer is very large, so the approximation of a weak perturbation no longer holds stringently. Second, if there is an overall linear background trend, the dominant slow Rabi dynamics would require long evolutions for us to see it, which are however numerically prohibitive.

We see from Eq.~\ref{eq:linear_response} that $\expval{\dot{N}_{3}(t)} \sim \sum_{k} A(k,\omega_{\text{rf}})$ and indeed the onset of the spectral response in $N_{3}(t)$, as extracted from our fits, agrees well with the lower onset of the spin-charge excitation continuum (grey dashed line) besides a broadening. The width of the resonance as observed here in $N_{3}$ is however much wider, compared to the momentum resolved spectra, due to the interplay of different momenta. In particular, the strong coupling to the excitations of `spin-wave' character (orange and red lines in Fig.~\ref{fig:nk3spectrum_weakU12_bothOmegaR}) gives rise to the long tail of the $N_{3}$ spectrum. Finally, we also show the (scaled) maximal transfer between $0 \leq Jt \leq 15\hbar$ for both driving amplitudes. It is peaked at the resonance, making clear that this is not only the driving frequency of greatest transfer rate, but also of overall net integrated transfer.

\section{Strongly attractive Hubbard model: response to weak driving}
\label{sec:strong_attraction}

\subsection{Momentum-resolved transfer to the third state}
We now turn to the case of strong interactions. We will perform the same detailed analysis of the upper state $\ket{3}$ populations for weak transfer $\hbar \Omega_{23} = 0.01J$, and compare the extracted transfer rates to exact calculations from Bethe ansatz.

\begin{figure}
  \includegraphics[width=1.0\columnwidth]{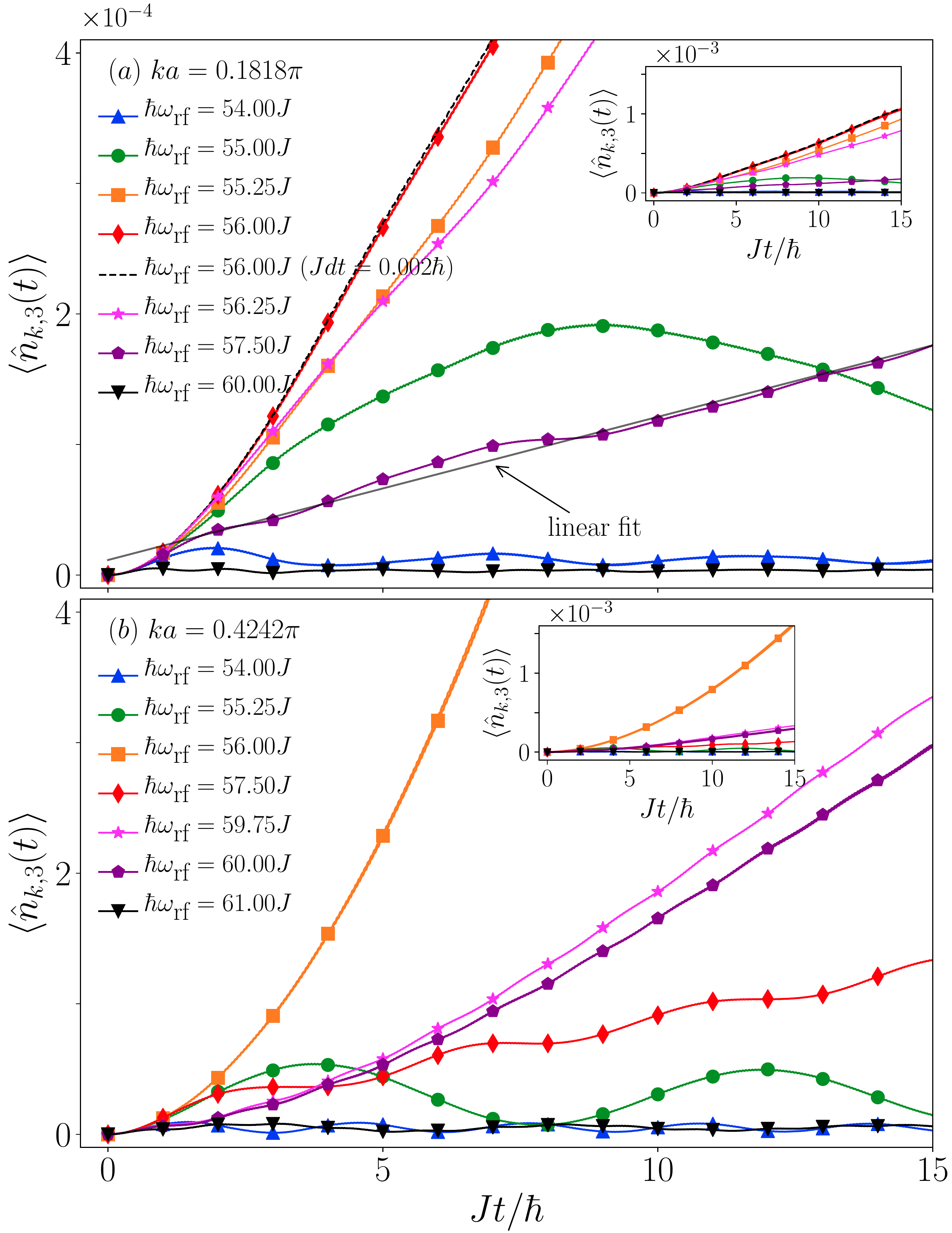}
  \caption{Time-dependence of the upper band population $\expval{\nop_{k,3}(t)}$ for selected momentum states at various driving frequencies $\omega_{\text{rf}}$ across the
          main resonance for a system of $L = 32$ sites at half-filling for interaction strength $U = -8J$. Level $|3\rangle$ is $\epsilon_3 = 50J$ in energy above levels $\ket{1}$ and $\ket{2}$, and the Rabi frequency is $\hbar \Omega_{23} = 0.01J$. The time-evolution can be separated into two regimes: a Rabi-like, and a linear response regime. Full grey lines are examples of linear fits, the extracted slopes are then used to construct the spectrum shown in Fig.~\ref{fig:nk3spectrum_strongU12_bothOmegaR_mA}. (a) $k a = 0.1818\pi$; (b) $k a = 0.4242\pi$.
          To ensure convergence of our results, we have separately varied the bond dimension ($D=400$), truncation error ($\epsilon = 10^{-13}$), and time step 
          ($Jdt = 0.002\hbar$) from the parameters given in section~\ref{sec:mps}. In (a) we show the convergence for $\hbar\omega_{\text{rf}}=54.0J,56.0J$, with the time step shown explicitly as a black dashed line and the remaining curves in the same colour as the data. In the lower panel (b) $\hbar\omega_{\text{rf}}=56.0J,60.0J$ convergence curves are overlaid in the same colours as the shown data. Only very small deviations are found.
          }
  \label{fig:nk3_strongU12_weakOmegaR}
\end{figure}

Examples of the evolution are shown in Fig.~\ref{fig:nk3_strongU12_weakOmegaR} for two momenta $ka = 0.1818\pi$ (a) and $ka = 0.4242\pi$ (b). For low $\omega_{\text{rf}} < 55J$ we see the dominant Rabi character in the evolution (fast oscillations with low transfer amplitude), however there is no consistent frequency beyond which we enter the linear region. This is a first indication that the spin-charge excitation continuum is strongly dispersive for large interactions. Indeed, the curve of largest transfer is found for increasing momentum at higher energies. For the presented momenta it is at $\hbar\omega_{\text{rf}} \sim 55.25J$ and $\hbar\omega_{\text{rf}} \sim 56.0J$, respectively (orange line in the two panels respectively). Driving the system close to maximal transfer, we see slow, large amplitude oscillations in its response. At nearby frequencies the evolution is strongly damped, while on resonance the frequency of oscillation is too slow for us to comment on the dephasing in this case. Once the frequency of the rf field can cause resonant excitations, the response is dominated by a net linear trend underlying the whole dynamics, with oscillations largely diminished. The transition into this linear regime occurs rather quickly. This is however not surprising, since we expect a stronger interaction induces level mixing and thereby a stronger coupling to a continuum.

\begin{figure} 
  \includegraphics[width=1.0\columnwidth]{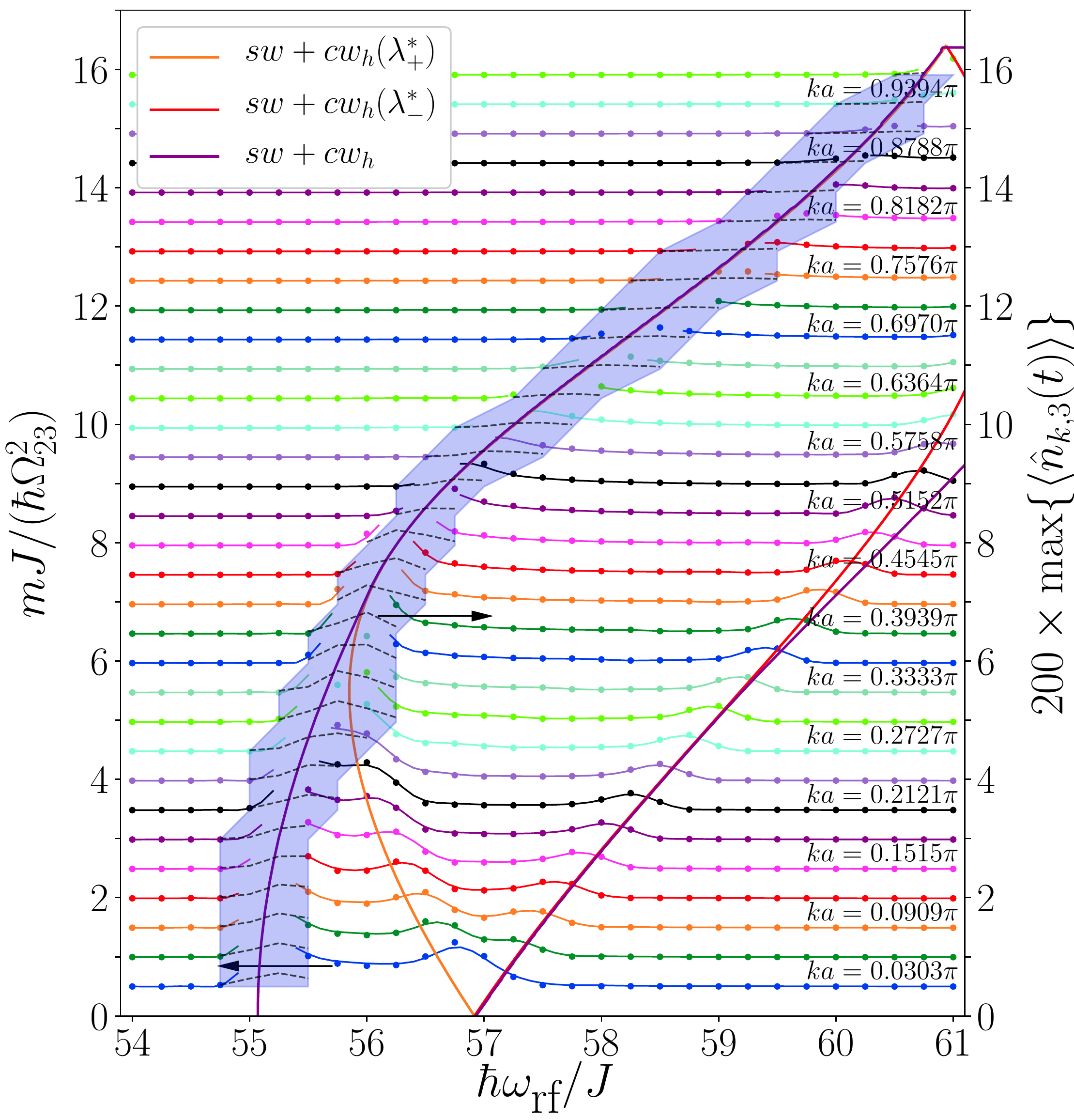}
  \caption{Rescaled momentum-resolved transfer rate to level $\ket{3}$, $m(k,\omega_{\text{rf}})/\Omega_{23}^{2}$, for a system of $L = 32$ sites at half-filling for 
          interaction strength $U = -8J$. Level $\ket{3}$ is $\epsilon_{3} = 50J$ in energy above levels $\ket{1}$ and $\ket{2}$. The dots represent the rescaled slopes for $\hbar \Omega_{23} = 0.01J$ and the lines, $\hbar\Omega_{23} = 0.1J$ (left axis, c.f. left arrow). These two data sets are found to be in good agreement. The shaded region corresponds to the frequency interval over which the time-evolution is Rabi-like. In this region, we report the maximum atom transfer in the time interval $0 \leq J t \leq 15\hbar$ (rigth axis, c.f. right arrow). The momentum values $ka = \frac{m\pi}{L+1}$ are equally spaced and are shifted vertically by $ka (L+1)/(2\pi) = m/2$, where m takes integer values. The bold solid lines are the lower and upper limits of the spin-charge continuum (purple) and two `spin-wave' excitations (orange and red lines) obtained from Bethe ansatz. 
          }
  \label{fig:nk3spectrum_strongU12_bothOmegaR_mA}
\end{figure}

As in the previous section, we analyze the curves in more detail by fitting a linear slope to the initial transient response ($m$), shown in Fig.~\ref{fig:nk3spectrum_strongU12_bothOmegaR_mA}. The general picture that emerged for $U=-2J$ holds here as well. The transfer is Rabi-dominated when driving below or above the spin-charge continuum. The shaded region close to some resonant excitations denotes evolutions we cannot fit linearly due to their slow frequency oscillations. We find that also for strong interactions the excitation lines are very clearly defined and in good agreement with the exact calculations from Bethe ansatz. We can clearly see the the dispersive spin-wave band joining the lower continuum edge at large momenta, which also coincides with the maximal momentum-resolved transfer in the upper level population. As for the response in general, the regime of non-zero effective transfer is given by the upper and lower edges of the excitation continuum. 

It is worth pointing out some differences in the spectral lines compared to the weakly interacting case. Firstly, we have fitted the $\expval{\nop_{k,3}}$ curves up to the other edge of the first Brillouin zone at $ka = \pi$. The reason lies in the broader momentum distribution of the ground state which leads to enhanced transfer also above the non-interacting Fermi momentum. This will be discussed in more detail in the following paragraph. Secondly, the curvature of the lower edge of the spin-charge continuum is much more pronounced which explains the widely differing resonance onsets in the momentum resolved curves for $\expval{\nop_{k,3}(t)}$ as we will detail below. Let us note that this also implies a very broad resonance peak in the $N_{3}$ spectrum, Fig.~\ref{fig:N3spectrum_strongU12_weakOmegaR}. In agreement with our findings for weak interactions we can confirm that this driving scheme strongly couples to the spin-wave degrees of freedom (red and orange line) in the system. Beyond the upper edge, we recover weak, oscillatory transfer, reminiscent of far-detuned Rabi oscillations. 

\begin{figure} 
  \includegraphics[width=1.0\columnwidth]{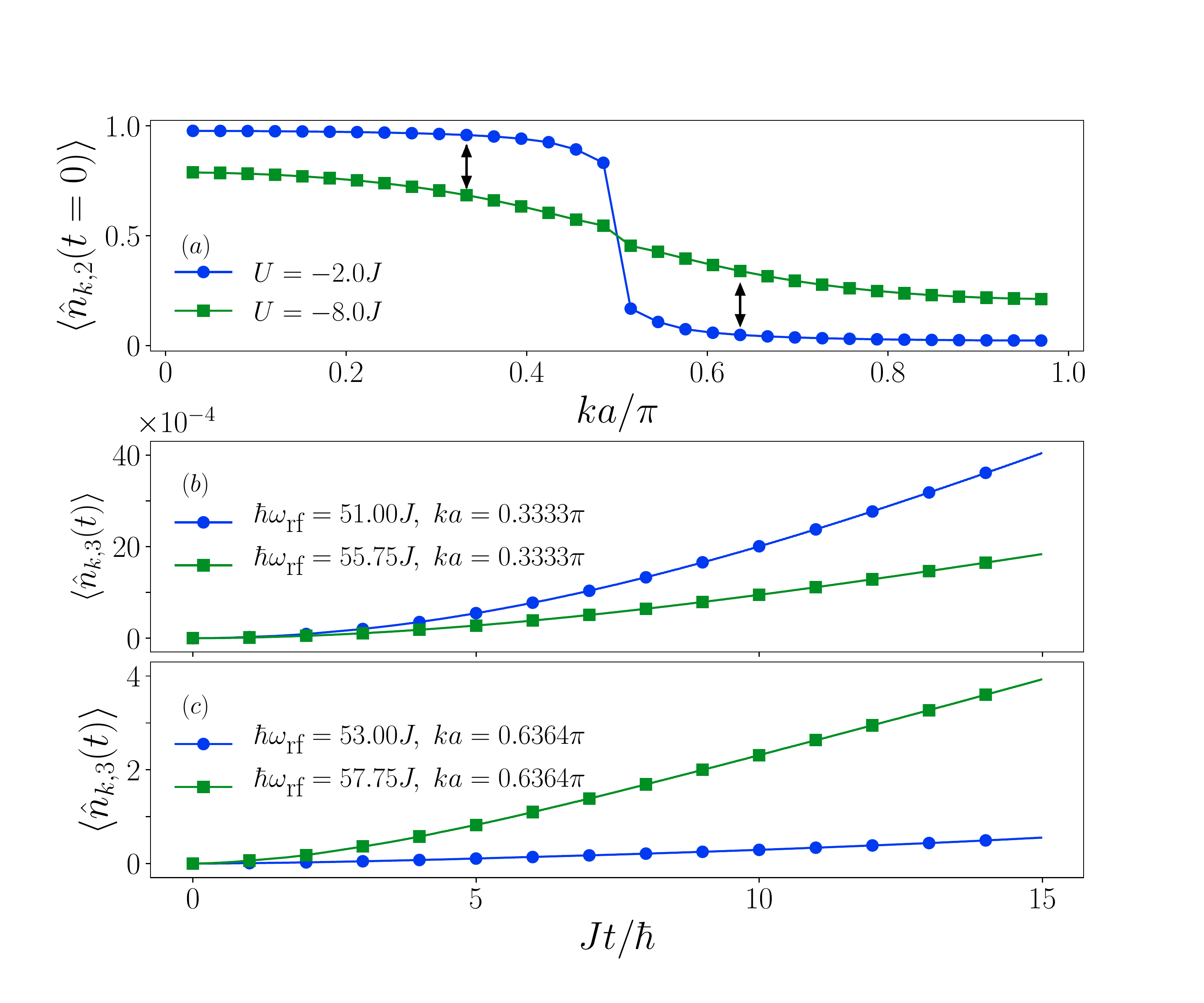}
  \caption{(a) Ground state distribution of $\expval{\nop_{k,2}(0)}$ for $U = -2J,-8J$. The two lower panels show the time-dependence of $\expval{\nop_{k,3}(t)}$ at
          (b) $k a = 0.3333\pi$ and  (c) $k a = 0.6364\pi$ near their respective resonances for a half-filled system of size $L = 32$ and $\hbar\Omega_{23} = 0.01J$.
          }
  \label{fig:GSnk2_nk3Comparison_bothU12}
\end{figure}

As we hinted above, the reason that we are able to extract a meaningful slope from the momentum-resolved evolution of the population of $\ket{3}$ for momenta above the Fermi momentum  $ka \geq \pi/2$ lies in its larger initial occupation, shown in Fig.~\ref{fig:GSnk2_nk3Comparison_bothU12} (a). Here we compare the initial ground state momentum distribution for the two interaction strengths considered. Whilst the $U=-2J$ distribution already shows a softening around the Fermi edge at $ka \sim \pi/2$ compared to the non-interacting Fermi-Dirac distribution, for strong interactions this edge is completely smeared out and rather resembles a slowly decaying function of momentum. For the marked momenta below ($ka = 0.333\pi$) and above ($ka = 0.6364\pi$) the Fermi surface, we plot the full time evolution in the lower two panels (b) and (c) respectively. Whilst for states below the Ferm edge the transfer is larger for $U = -2J$ (b), the situation is reversed above the Fermi edge (c). This corroborates our assertion, that it is the very different occupation of $\expval{\nop_{k,2}(t=0)}$ in the different regions of the Brillouin zone, that affects the observed transfer to the upper level.

\subsection{Evolution of the momentum distribution}
\begin{figure} 
  \includegraphics[width=1.0\columnwidth]{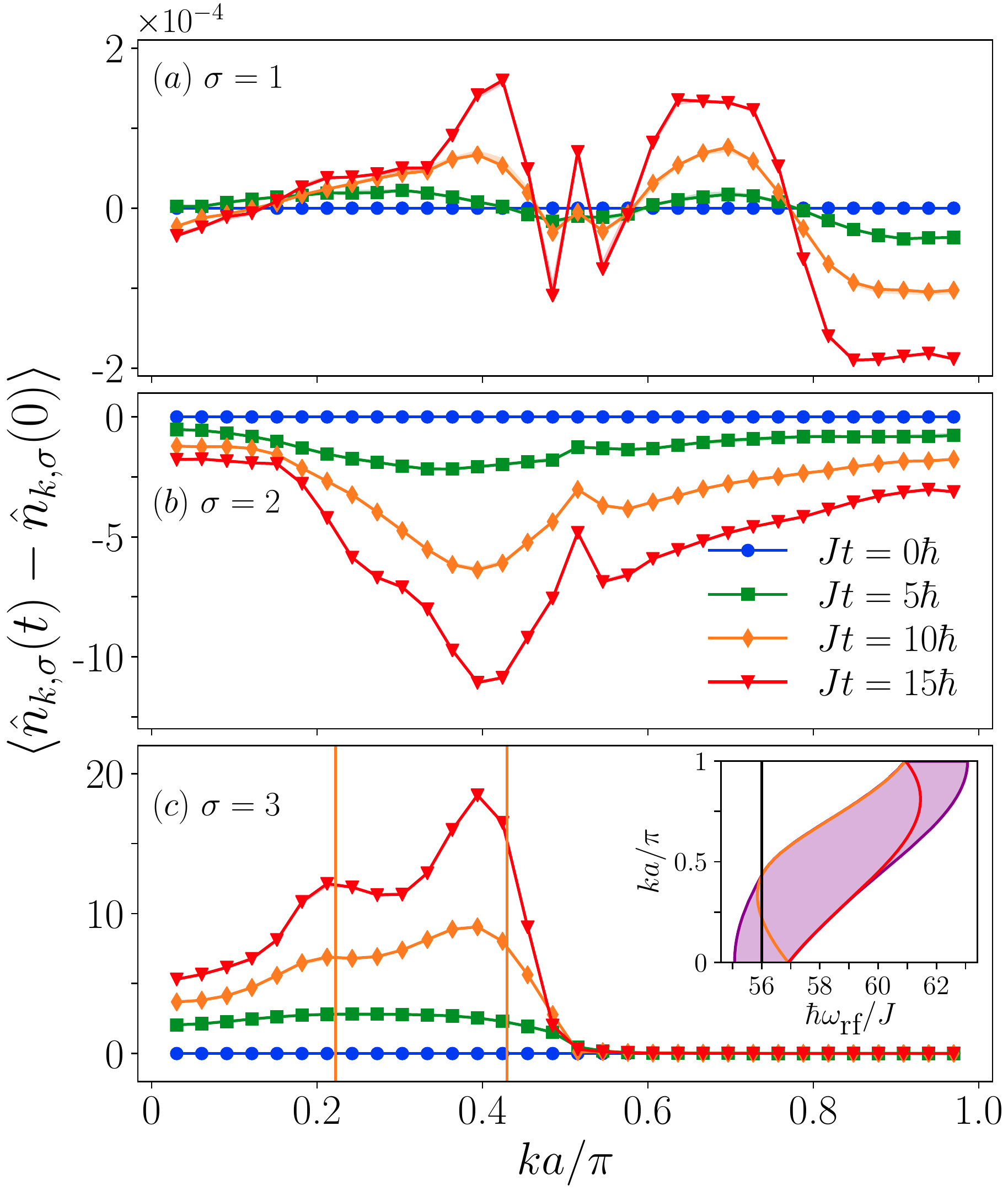}
  \caption{Evolution of the differences of the momentum distributions at times $t$ compared to the initial time, $\expval{\nop_{k,\sigma}(t)}-\expval{\nop_{k,\sigma}(0)}$,
          for $U = -8J$, $\hbar \Omega_{23} = 0.01J$, and $\hbar \omega_{\text{rf}} = 56.0J$ for $\sigma=\{1,2,3\}$ in panels (a-c) respectively. We show the different times as marked in legend in (b). The inset in (c) shows the spin-charge excitation continuum (purple region bounded by purple lines), along with two particular excitations of `spin-wave' character (orange and red lines). The black vertical line marks the driving frequency $\omega_{\textrm{rf}}$ (inset), while the vertical, orange lines in the lower panel (c) mark the momenta, at which the driving frequency $\omega_{\textrm{rf}}$ is resonant with the `spin-wave' excitation (orange line, inset). The shown data was obtained for a bond dimension $D=500$, truncation error $\epsilon=10^{-12}$, and time step $Jdt = 0.001\hbar$. 
          To ensure convergence of our results, we have separately varied the bond dimension ($D=600$), truncation error ($\epsilon = 10^{-13}$), and time step ($Jdt = 0.0005\hbar$). The maximal and minimal deviation is shown as a shaded region around the corresponding curve (same colour respectively).
          }
  \label{fig:FermiEdge_strongU12_weakOmegaR_res}
\end{figure}

\begin{figure} 
  \includegraphics[width=1.0\columnwidth]{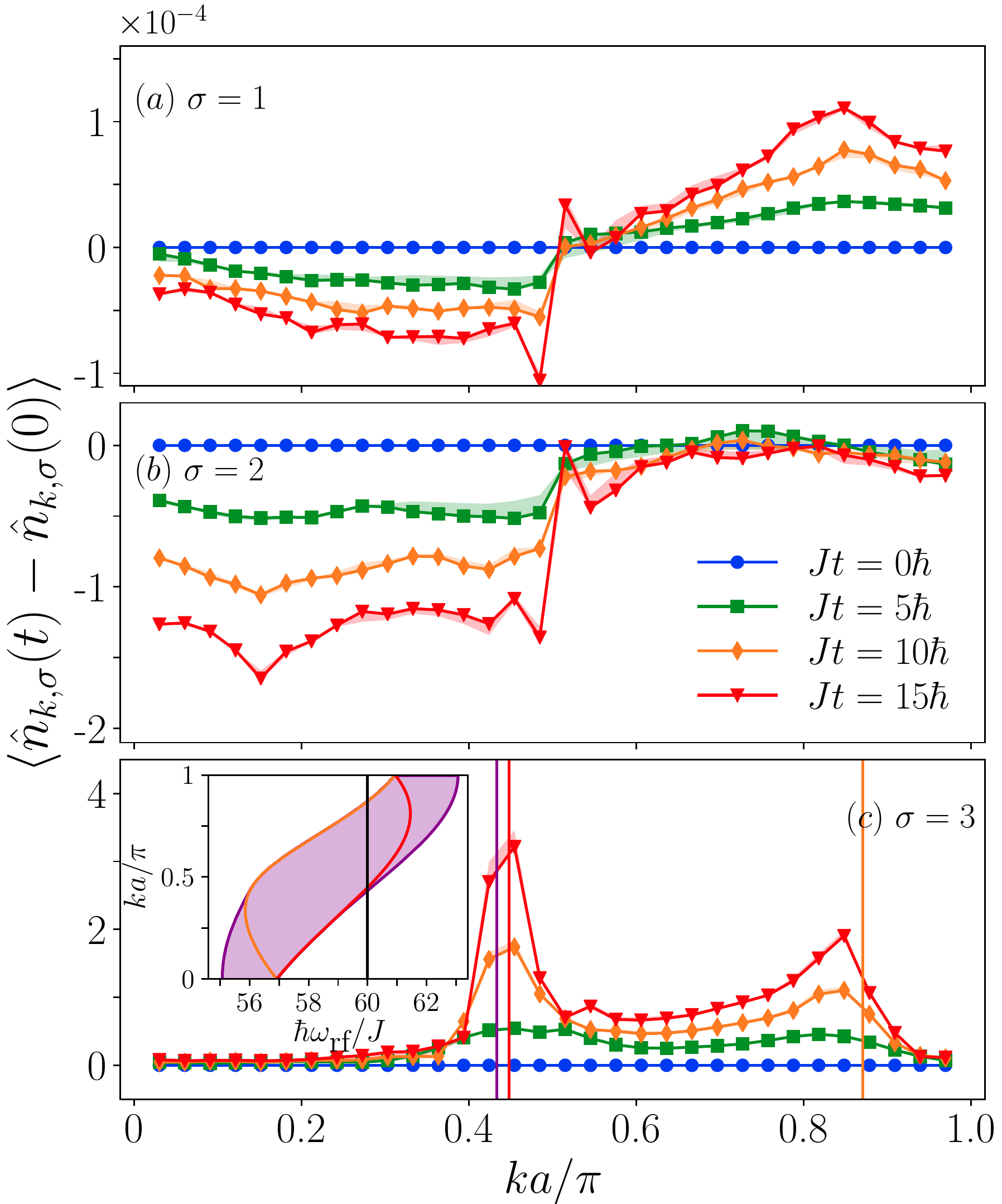}
  \caption{Evolution of the differences of the momentum distributions at times $t$ compared to the initial time, $\expval{\nop_{k,\sigma}(t)}-\expval{\nop_{k,\sigma}(0)}$,
          for $U = -8J$, $\hbar \Omega_{23} = 0.01J$, and $\hbar \omega_{\text{rf}} = 60.0J$ for $\sigma=\{1,2,3\}$ in panels (a-c) respectively. 
          We show the different times as marked in legend in (b). The inset in (c) shows the spin-charge excitation continuum (purple region bounded by purple lines), along with two particular excitations of `spin-wave' character (orange and red lines). The black vertical line marks the driving frequency $\omega_{\textrm{rf}}$ (inset), while the coloured vertical lines in the lower panel (c) mark the momenta, at which the driving frequency $\omega_{\textrm{rf}}$ is resonant with either the `spin-wave' excitations (red and orange), or the upper spin-charge continuum edge (purple). The shown data was obtained for a bond dimension $D=500$, truncation error $\epsilon=10^{-12}$, and time step $Jdt = 0.001\hbar$. 
          To ensure convergence of our results, we have separately varied the bond dimension ($D=600$), truncation error ($\epsilon = 10^{-13}$), and time step ($Jdt = 0.0005\hbar$). The maximal and minimal deviation is shown as a shaded region around the corresponding curve (same colour respectively).
          }
  \label{fig:FermiEdge_strongU12_weakOmegaR_blue}
\end{figure}

We now turn to the discussion of the evolution of the momentum distributions $\expval{\nop_{k,\sigma}}$, and will focus on two driving frequencies, close to the resonance of maximal integrated transfer at $\omega_{\text{rf}} = 56J$ (Fig.~\ref{fig:FermiEdge_strongU12_weakOmegaR_res}), and resonant driving in the upper half of the Brillouin zone, above the resonance in $N_{3}$ (Fig.~\ref{fig:FermiEdge_strongU12_weakOmegaR_blue}). 

Following on from our discussions in the weakly interacting system, we observe the appearance of the characteristic two-peak structure in the lowest panel (c) for state
$\ket{3}$ of Fig.~\ref{fig:FermiEdge_strongU12_weakOmegaR_res}. The vertical lines mark the momenta to which the rf drive is coupling resonantly and we find them to be in very good agreement with the enhanced transfer. Initially the rf drive depletes $\ket{2}$ for a broad range of momenta, but eventually the resonant coupling to momentum states $ka \sim 0.2\pi$ and $ka \sim 0.4\pi$ becomes the dominant transfer mechanism. This can be seen in the arising dip structure for example at long times $Jt \gtrsim 10\hbar$. The opposite in turn holds for the population gain in level $\ket{3}$. Contrary to weak interactions however, the larger momenta get also significantly depleted ($\sigma=2$), but are \emph{not} in the same way populated into the corresponding momentum states for $\sigma=3$. This strongly supports our interpretation, that the transfer to the upper level is largely going through the resonant momentum channels arising from the coupling to the spin-wave excitations (orange and red lines in the inset of (c)). Meanwhile the population of $\ket{1}$ shows population redistribution across all momenta. Since the dynamics in $\ket{1}$ is purely induced by the interaction $U$, it is not surprising that the effect is seen more strongly here. Finally, it is important to point out that the overall transfer in all levels is significantly reduced compared to the weaker interaction, by nearly an order of magnitude. We argue that the strong interaction leads to an increased rate of dephasing, and thus reduces coherent transfer.

For large driving frequencies, Fig.~\ref{fig:FermiEdge_strongU12_weakOmegaR_blue}, the two peaks move into the upper half of the Brillouin zone, where states $ka \sim 0.4\pi$ and $ka \sim 0.9\pi$ are driven resonantly. The $\expval{n_{k,3}}$ evolution (c) monotonically increases with time, predominantly at the resonant momentum states (marked by the vertical lines), oscillations are damped out, and the strong interactions place the drive inside the linear regime. Whilst $\expval{\nop_{k,3}}$ increases strongly at the zone centre and upper edge (c), $\expval{\nop_{k,2}}$ does \emph{not} show the complimentary depletion. Instead it is mainly emptied for all momenta $k \leq k_{F}$ (b). This points to a strong redistribution of the particles, as confirmed by panel (a). Particles are moved from below to above the Fermi surface and the system is heated in the process.

\subsection{Evolution of the pair distribution}
\begin{figure} 
  \includegraphics[width=1.0\columnwidth]{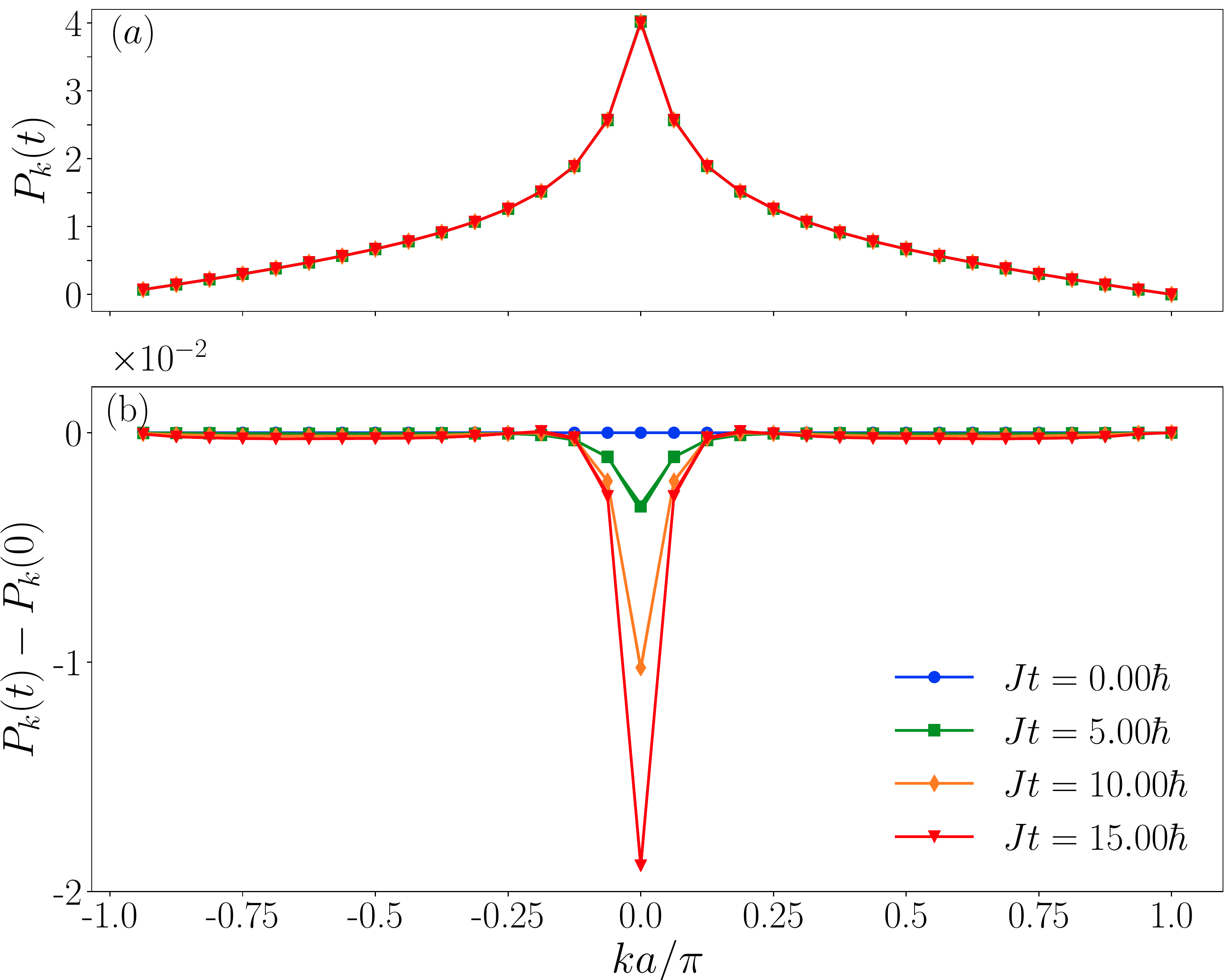}
  \caption{Evolution of the pair distribution $P_{k}(t)$ for the attractive Hubbard model of size $L = 32$ at half-filling for $U = -8J$, $\hbar \Omega_{23} = 0.01J$ and
          $\hbar \omega_\text{rf} = 56.0J$. 
          To ensure convergence of our results, we have separately varied the bond dimension ($D=400$), truncation error ($\epsilon = 10^{-13}$), and time step 
          ($Jdt = 0.002\hbar$) from the parameters given in section~\ref{sec:mps}. These are shown in the same colour as the data, the numerical error is therefore below the linewidth.
          }
  \label{fig:PairDistribution_strongU12_weakOmegaR_res}
\end{figure}

Here we briefly comment on the evolution of the pair distribution of Eq.~\ref{eq:pair_distribution} for strong interactions and maximal net transfer $\hbar\omega_{\text{rf}} = 56.0J$. The large transfer is reflected in the pair correlation as a monotonic depletion of the pairs close to $ka \sim 0$, Fig.~\ref{fig:PairDistribution_strongU12_weakOmegaR_res}. In contrast to the weaker interaction, Fig.~\ref{fig:PairDistribution_weakU12_weakOmegaR_res}, here pairs are tightly bound together on a site. This seems to lead to a greater stability of short range pair coherence, compared to the case of $U=-2J$.

\subsection{Total transfer to the third state}
\begin{figure} 
  \includegraphics[width=1.0\columnwidth]{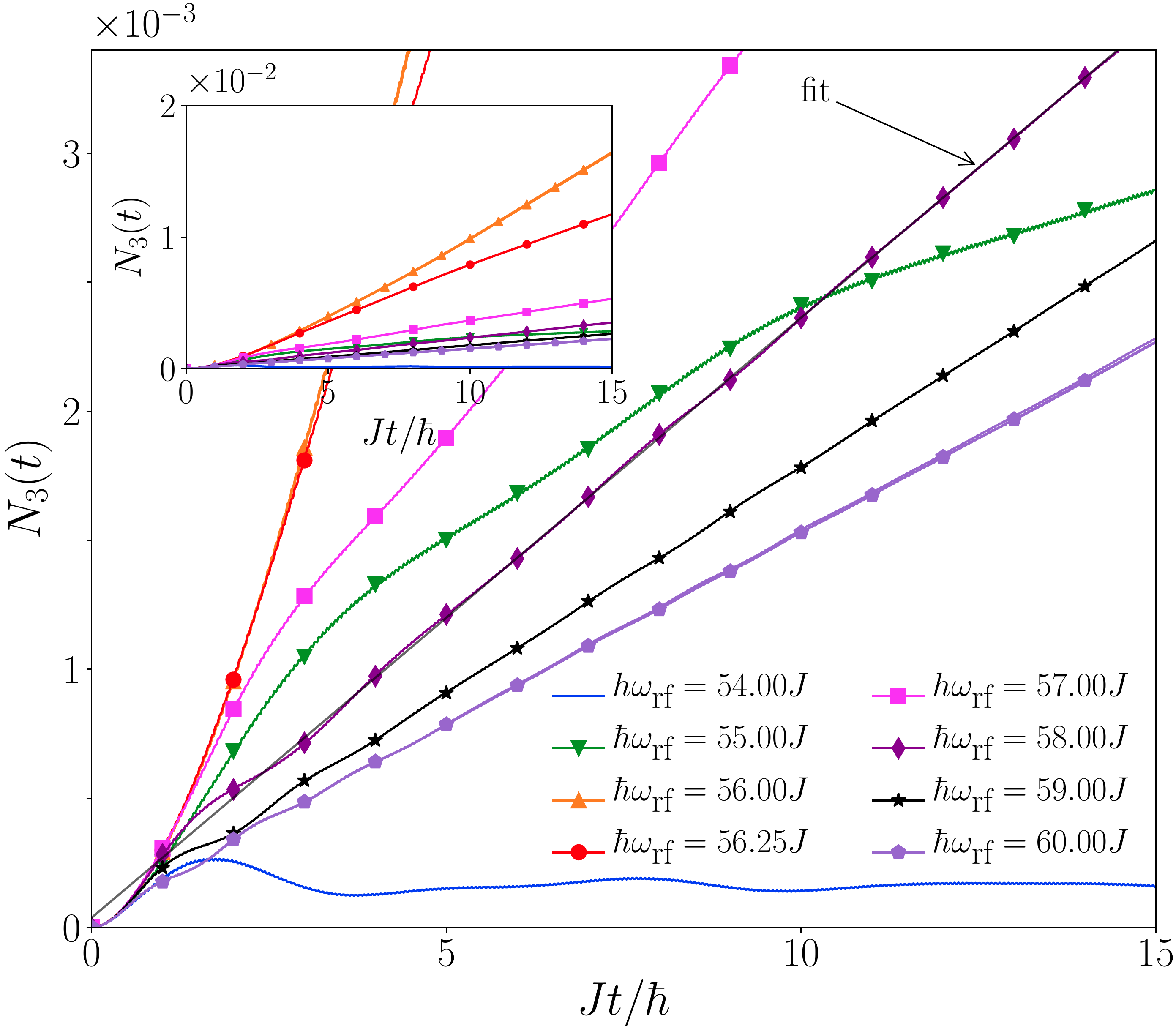}
  \caption{Time-dependence of the total transfer to the third state, $N_{3}(t)$, for the attractive Hubbard model of size $L=32$ at half-filling for $U = -8J$,
          $\hbar\Omega_{23} = 0.01J$ and several driving frequencies $\omega_{\textrm{rf}}$. The main plot focuses on the curves for off-resonance driving, whilst the inset puts these in context when compared to resonantly driven situations. The grey solid line is the linear portion of a fit to the time-dependence for $\hbar\omega_\text{rf} = 60.0J$. 
          To ensure convergence of our results, we have separately varied the bond dimension ($D=400$), truncation error ($\epsilon = 10^{-13}$), and time step 
          ($Jdt = 0.002\hbar$) from the parameters given in section~\ref{sec:mps}. These are shown in the same colour as the data, the simulation error is therefore below the linewidth.
          }
  \label{fig:N3_strongU12_smallOmegaR}
\end{figure}

We conclude our discussion of the influence of strong interactions on the rf drive by looking at the experimentally most accessible quantity, the total upper level population
$N_{3}(t)$, shown in Fig.~\ref{fig:N3_strongU12_smallOmegaR} for various driving frequencies. For red-detuned driving frequencies below the continuum edge,
$\hbar \omega_{\text{rf}} \lesssim 55J$, the integrated transfer oscillates around a small long time value. Beyond this driving frequency, the oscillatory behaviour gradually goes over into a linear rise. We do not observe a relatively sharp onset of the linear regime as was the case for weak interactions, since the lower edge of the excitation continuum is curved more strongly as we mentioned above (c.f. Fig.~\ref{fig:nk3spectrum_strongU12_bothOmegaR_mA}). For $\hbar \omega_{\text{rf}} \sim 56J$ the net transfer is maximal. Strong interactions lead to an enhanced dephasing, and as a result the oscillations on top of the linear increase are strongly damped out, or not observable at all. To access the spectrum of $N_{3}$, we fit this region with a linear function, as exemplified by the grey solid line in Fig.~\ref{fig:N3_strongU12_smallOmegaR}.

\begin{figure} 
  \includegraphics[width=1.0\columnwidth]{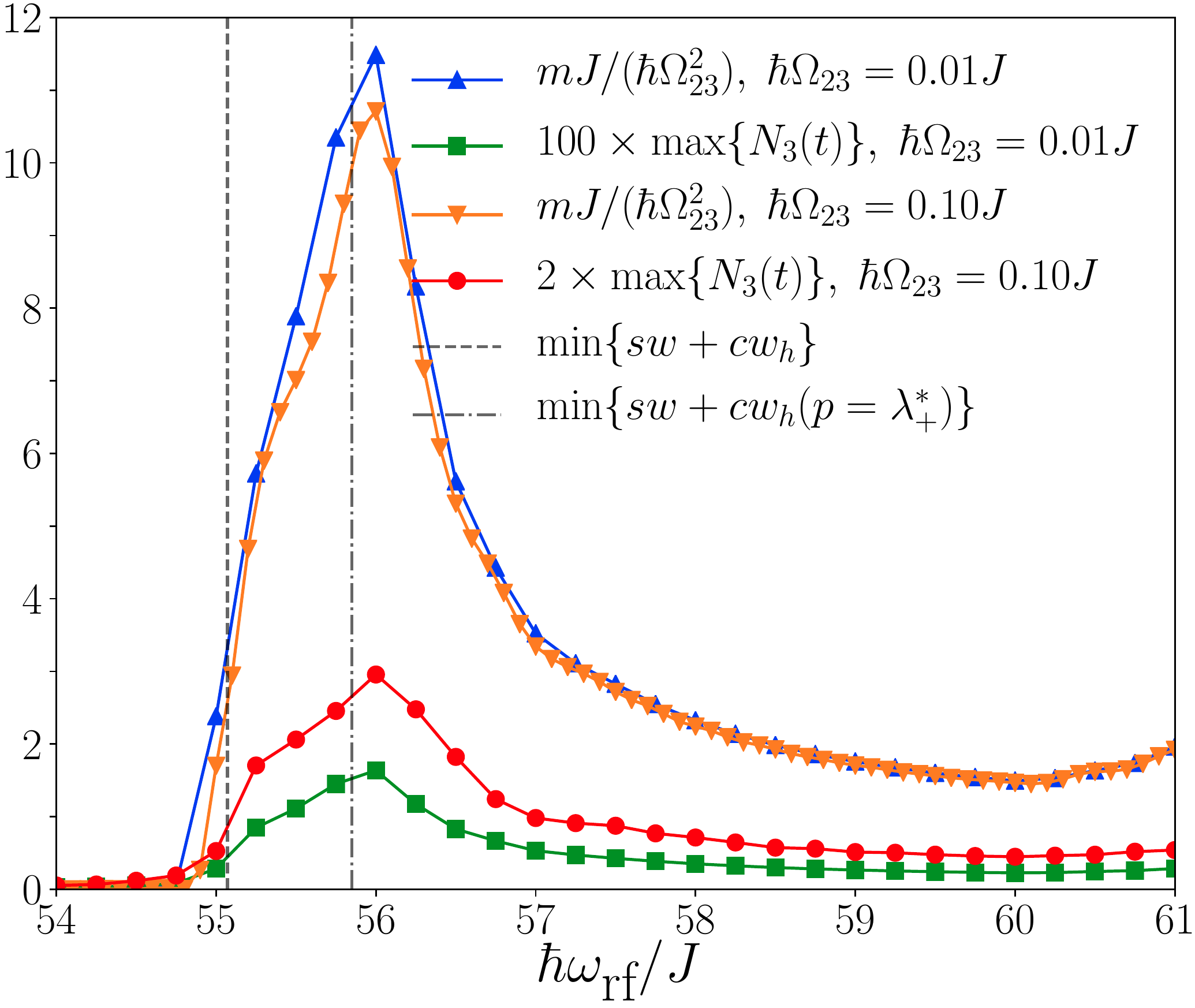}
  \caption{Radiofrequency transfer rate (scaled by the squared Rabi frequency) and scaled maximum net transfer between $0 \leq Jt \leq 15\hbar$ for a system of size $L = 32$,
          $U = -8J$, and  $\hbar\Omega_{23} = 0.01J$ and $0.1J$. The grey dashed line marks the lower edge of the spin-charge excitation continuum, whilst the grey dash-dotted line highlights the minimum energy of the `spin-wave'-type excitations, $\epsilon_{sw}(q) + \epsilon_{cw}(\lambda^{*}_{+})$, where $p^{h}_{cw}(\lambda^{*}_{+}) = \pi/(2a)$.
          }
  \label{fig:N3spectrum_strongU12_weakOmegaR}
\end{figure}

The spectrum, shown in Fig.~\ref{fig:N3spectrum_strongU12_weakOmegaR}, is peaked around $\hbar \omega_{\text{rf}} = 56.0J$ and is much wider than in the attractive case. As detailed when discussing the momentum resolved spectra, the width originates from the strong curvature of the lower excitation band as seen in the single particle spectral function, Fig.~\ref{fig:nk3spectrum_strongU12_bothOmegaR_mA} (orange line). For clarity we have marked the lower onset of the spin-charge continuum (grey dashed line) as well as the minimum energy of this particular excitation with `spin-wave' character (grey dash-dotted line). Whilst the former sets the \emph{onset} of the spectral response of $N_{3}$, the latter dictates its resonance position, i.e. the frequency where the integrated transfer is maximal. We report an overall very good agreement with the spectral features obtained from our fitting procedure. The resonance position is shifted from the non-interacting result of $\hbar \omega_{\text{rf}} = 50J$ to significantly higher energies due to the interaction. It is important to note that here we are able to perform our fitting analysis throughout and across the resonance region because the strong interaction allows for sufficient mixing of the levels already at the lower edge of the excitation continuum. For far red-detuned drivings, the response is still oscillatory and Rabi-like (very weak transfer), but closer to the resonance, scattering and interactions obscure this picture and give rise to saturation (maximal transfers of up to $60\%$) and non-oscillatory behaviour, which gradually gives way to the linear response regime as shown in Fig.~\ref{fig:N3_strongU12_smallOmegaR}.

\section{Conclusion}
\label{sec:conclusion}
We studied in this work the response to rf driving of a system described by the half-filled attractively interacting one-dimensional Fermi-Hubbard model using the time-dependent matrix product state algorithm. The rf field drives the system away from equilibrium by inducing particle transfer to a free, upper band, whose population is monitored in time. The evolution explores two different dynamical regimes, one with a strong Rabi-character and another in the linear regime. While the former exhibits (off)resonant many-body oscillations in the upper level population, the latter emerges when the drive couples to a continuum of states. Interestingly, even though the driven system is not always in the linear response regime, we are still able to extract the underlying spectra to a reasonable accuracy. 

Our numerical simulations allow us to access the complete time-evolution of the system throughout the drive, where we observe complex, intricate dynamics. Many features of the extracted spectra are in good agreement with exact Bethe ansatz calculations, and particularly the momentum-resolved upper level population $\expval{\nop_{k,3}(t)}$ provides great insight into the underlying excitation structure of the system, and the way the rf drive couples to these excitations. As such, rf spectroscopy is an invaluable tool to probe the system as it offers direct access to the single-particle spectral function in the weak-coupling regime. In addition, we showed that this rf technique can be employed to investigate many-body coupling mechanisms away from equilibrium. This was done by monitoring the momentum density redistributions and the evolution of the pair correlations during the drive. Considering such non-equilibrium physics goes beyond a linear response treatment and was achieved here by conducting time-dependent matrix product state simulations.

Our present work has explored the intricate nature of the rf transfer and given a detailed account of its potential to study atomic gases in experimentally realistic settings. Due to the generality of our model, these discussions are relevant both to rf spectroscopy studies, but are also amenable to investigations of multi-orbital, interacting quantum many-body systems~\cite{WernerEckstein2018,RinconFeiguin2018}. As a future direction, our results could be analyzed from another angle by focusing on the motion of defects created by the rf transfer.

\section{Acknowledgements}
We acknowledge useful discussions with T. Giamarchi and C. Salomon, and funding from the German Research Foundation (DFG) under project number 277146847 - CRC 1238 (C05), project number 277625399 - TRR 185 (B4) and under Germany Excellence Strategy Cluster of Excellence Matter and Light for Quantum Computing (ML4Q) EXC 2004/1 390534769, from the European Research Council (ERC) under the Horizon 2020 research and innovation programme, grant agreement No. 648166 (Phonton) and from the Natural Sciences and Engineering Research Council of Canada (NSERC).

%\bibliographystyle{apsrev4-1}  
%\bibliography{bib,references}
\bibliography{RFspectroscopy}
\end{document}